\documentclass[prb,twocolumn,amsmath,amssymb,eqsecnum,floatfix]{revtex4-2}
\usepackage{epsfig, graphicx,graphics,amsmath,amssymb,float}
\usepackage[T1]{fontenc}
\usepackage[latin9]{inputenc}
\usepackage{amsmath}
\usepackage{amssymb}
\usepackage{appendix}
\usepackage{amscd}
\usepackage{bm}
\usepackage{psfrag}
\usepackage{bbm} 
\usepackage{babel}
\usepackage{wasysym}
\usepackage{mathrsfs}
\usepackage{color}
\usepackage[normalem]{ulem}
\usepackage{tabularx} 
\usepackage{tikz}
\usetikzlibrary{calc,decorations.markings}
\usepackage[final]{hyperref} 
\hypersetup{
	colorlinks=true,       
	linkcolor=blue,        
	citecolor=blue,        
	filecolor=magenta,     
	urlcolor=blue         
}

\newcommand{\new}[1]{\textcolor{black}{#1}}

\newcommand{\br}{{\bf r}}
\newcommand{\bk}{{\bf k}}

\usepackage{mathtools}

\DeclarePairedDelimiter\ket{\lvert}{\rangle}
\DeclarePairedDelimiterX\braket[2]{\langle}{\rangle}{#1 \delimsize\vert #2}

\begin{document}

\title{Green's Function Approach to Josephson Dot Dynamics and Application to Quantum Mpemba Effects }

\author{Kateryna Zatsarynna}
\affiliation{Institut f\"ur Theoretische Physik, Heinrich-Heine-Universit\"at, D-40225  D\"usseldorf, Germany}
\author{Andrea Nava}
\affiliation{Institut f\"ur Theoretische Physik, Heinrich-Heine-Universit\"at, D-40225  D\"usseldorf, Germany}
\author{Reinhold Egger}
\affiliation{Institut f\"ur Theoretische Physik, Heinrich-Heine-Universit\"at, D-40225  D\"usseldorf, Germany}
\author{Alex Zazunov}
\affiliation{Institut f\"ur Theoretische Physik, Heinrich-Heine-Universit\"at, D-40225  D\"usseldorf, Germany}

\begin{abstract}
 We develop a Green's function approach for the nonequilibrium dynamics of multi-level
 quantum dots coupled to multiple fermionic reservoirs in the presence of a bosonic environment.
 Our theory is simpler than the Keldysh approach and goes beyond  scattering state constructions.
 In concrete terms, we study 
 Josephson junctions containing a quantum dot and coupled to an electromagnetic environment.
 In the dot region, spin-orbit interactions, a Zeeman field, and in principle also Coulomb interactions can be included.
 We then study quantum Mpemba effects, assuming that the average phase difference across the Josephson junction is subject to a rapid quench.
 For a short single-channel junction, we show that both types of quantum Mpemba effects allowed in open quantum systems are possible. 
 We also study an intermediate-length junction, where spin-orbit interactions and a Zeeman field are included. Again quantum Mpemba effects are predicted. 
\end{abstract}
\maketitle

\section{Introduction} \label{sec1}

In many-body quantum systems, low-energy excitations, in particular of fermionic type, are not always easy to find by solving an eigenvalue problem based on the full microscopic Hamiltonian. An often more efficient alternative is to extract them by tracing out high-energy degrees of freedom.  For instance, 
within the Green's function (GF) approach, quasiparticles are identified through the poles of the single-particle GF \cite{Nazarov2009}. Alternatively, the low-energy theory can be formulated in terms of an effective action, generally with a time-nonlocal Lagrangian. Then, in the presence of external forces and/or dissipative processes, the nonequilibrium dynamics of quasiparticles can be addressed by the celebrated Lindblad master equation \cite{breuer2007theory}. This equation requires knowledge of the quasiparticle transition rates for all scattering events caused by external perturbations, including those induced by the dissipative environment. However, it is generally difficult to calculate these rates if the quasiparticles obey an equation of motion (EOM) that is nonlocal in time and cannot be described within the canonical Hamiltonian formalism. Specifically, Fermi's golden rule cannot be applied directly  in such cases.

A prototypical example for such systems are quantum dot Josephson junctions, referred to as ``Josephson dots'' from now on, which we study in this paper. However, it should be clear that the concepts put forward in Sec.~\ref{sec2}  are generally applicable to arbitrary quantum systems coupled to multiple  reservoirs.
In the phase-biased regime, the low-energy fermionic modes of a Josephson dot are represented by Andreev bound states (ABSs), with phase-dependent energies below the \new{Bardeen-Cooper-Schrieffer (BCS)} superconducting pairing gap $\Delta$ in the leads (assumed to be identical in all leads for simplicity).  These states are localized near the junction, which in our model contains a quantum dot, and carry a supercurrent. They are thus inherently linked to the phase dynamics. To calculate transition rates pertaining to ABSs in the presence of phase fluctuations induced by environmental fluctuations, a standard but somewhat cumbersome method \cite{Nazarov2009,Giuliano_2013,Giuliano_2014,Zazunov2014,Campagnano_2015,Park2017,Minutillo2018,Ackermann2023,Zatsarynna2024,Fauvel2024,Lidal2024}
is to solve the Bogoliubov-de Gennes (BdG) equations in order to find the full wave functions for all quasiparticle states, including those for continuum fermions with energy above $\Delta$. 

Below we introduce and apply a simpler, and arguably more elegant, strategy based on a GF approach to solve the Heisenberg EOMs for the fermionic fields. After averaging over lead fermions, we show how to handle the time-nonlocal and, in general, non-unitary dynamics of Andreev states and calculate all inter-level transition rates, 
including transitions to the continuum.   We allow for spin-orbit interaction (SOI), a Zeeman field, and electron-electron interactions in the quantum dot region. 
Our approach offers several key advantages compared to established alternative theoretical methods:
(i) It avoids the complexity of a full-fledged Keldysh approach \cite{Nazarov2009,Weiss_2012} yet allows to treat nonequilibrium problems.  
(ii) It avoids double counting problems by construction.  Such subtleties often occur for superconducting models in the presence of SOI \cite{Hasan2010}. 
(iii) There is no need for an explicit calculation of BdG eigenstates in order to compute transition rates entering the Lindblad equation governing the ABS dynamics. 
(iv) In contrast to BdG-based schemes, our approach also allows to account for Coulomb interactions in a natural manner, even though the application in Sec.~\ref{sec3} assumes a noninteracting dot.

In Sec.~\ref{sec3}, we illustrate our GF-based formalism to a study of the quantum Mpemba effect (QME) for a spin-orbit coupled Josephson dot subject to a magnetic Zeeman field in the 
presence of an electromagnetic environment. 
Quantum generalizations of the classical thermal Mpemba effect \cite{Mpemba_1969,Jeng2006,Lu2017,Lasanta2017,Jesi2019,Torrente2019,Santos2020,Megias2022,Chetrite2021,Walker2023}
have garnered a lot of recent attention \cite{Murciano_2024,Rylands2023,Shapira2024,Joshi2024,Chatterjee2023,Chatterjee2023_2,Wang2024,Strachan2024,liu2024,Turkeshi2024,Kochsiek2022,Klich2019,Ares2024,Nava2024,Wang2025,Dong2024,Qian2025,Ares_2025,Teza_2025}.  
The classical Mpemba effect   occurs if two system copies are prepared in equilibrium states at different temperatures $T_h$ (``hot'') and $T_c$ (``cold''), respectively. For each copy, one performs a sudden temperature quench at time $t=0$ to the final temperature $T_{\rm eq}$ with $T_{\rm eq}<T_c<T_h$.  Given the corresponding relaxation times $\tau(T_{h/c})$ towards the final equilibrium state for temperature $T_{\rm eq}$ reached for $t\to \infty$, the classical Mpemba effect takes place if shortcut pathways in the effective energy landscape exist  such that  $\tau(T_h)<\tau(T_c)$  \cite{Ibanez2024,Adalid2024}. 
In many cases, $\tau$ can be estimated as the time the system needs to undergo a phase transition \cite{Mpemba_1969,Nava2019,Vadakkayil2021,Pemartin_2021}. For systems without phase transitions,  
$\tau$  instead has to be extracted from a properly chosen monitoring function \cite{Lu2017}.

Different quantum generalizations of the classical Mpemba effect have appeared recently, 
with first experiments already available for trapped ions \cite{Shapira2024,Joshi2024}.
The theory of the QME in \emph{closed} quantum systems was addressed, e.g., in Refs.~\cite{Murciano_2024,Rylands2023,Turkeshi2024}.
For \emph{open nonequilibrium} quantum systems coupled to multiple reservoirs (``baths'') \cite{Weiss_2012,breuer2007theory}, 
the competition between stochastic relaxation processes and 
quantum effects
may drive the system towards a (nonequilibrium or equilibrium) stationary state, denoted by 
(N)ESS below.
Also under such circumstances, one can encounter a QME which is characterized and sometimes 
even dominated by quantum correlations, entanglement, and quantum coherence. The practical importance of the QME comes from the possibility to dramatically accelerate certain processes and transitions.

Below we adapt the general protocol for identifying the QME proposed by two of us \cite{Nava2024} to the Josephson dot problem.  
This protocol can unambiguously identify the QME in open nonequilibrium systems with Markovian dynamics, where two types of QME are possible in general, see Sec.~\ref{sec3} for details.
We consider a quench of the average phase difference $\phi_0$  across the Josephson junction. Experimentally, this can be achieved, e.g., by quenching a magnetic flux \cite{Nazarov2009}. We compare two copies of the system corresponding to pre-quench values $\phi_0^{(c)}$ and $\phi_0^{(f)}$, which are respectively quenched at time $t=0$ to the same post-quench value $\phi_0^{({\rm eq})}$ describing the final  configuration.  The superscripts $(c)$ and $(f)$ refer to ``close'' vs ``far'' with respect to the final phase difference, corresponding to the ``cold'' vs ``hot'' case in the classical Mpemba effect, i.e.,
we require $\left|\phi_0^{(c)}-\phi_0^{({\rm eq})}\right|<\left|\phi_0^{(f)}-\phi_0^{({\rm eq})}\right|$.

Following Ref.~\cite{Nava2024}, in order to monitor the distance of the post-quench quantum state $\rho_A(t)$, which describes the ABS sector, from the final stationary state $\rho_{A,{\rm stat}}$ reached for $t\to \infty$,
we employ the trace distance \cite{Nielsen2000},
\begin{equation}\label{tracedist}
\mathcal{D}_T (\rho_A (t))=\frac12\mathrm{Tr}\left|\rho_A(t)-\rho_{A,{\rm stat}}\right|,
\end{equation}
As discussed in Refs.~\cite{Lu2017,Nava2024}, one may equivalently choose a different distance function 
as long as ${\cal D}_T(\rho_A(t))$ is a monotonically non-increasing, continuous, and convex function of time. 
The trace distance satisfies these consistency relations under Markovian dynamics \cite{Nielsen2000,Wang2009}.
In addition, for the application studied here, it could be measured experimentally in terms of microwave spectroscopy, see Sec.~\ref{sec3}.
As we show below, $\rho_A(t)$ obeys a Markovian Lindblad equation.

The remainder of this paper is structured as follows. In Sec.~\ref{sec2}, we describe the
GF formalism as applied to the Josephson dot problem.  In Sec.~\ref{sec3}, we then study the 
emergence of the QME in this system, where we make concrete predictions for parameter regimes
where QMEs are expected.  Finally, we briefly conclude in Sec.~\ref{sec4}.  The appendix contains technical
details on the calculation of wave functions for the central quantum dot defining the junction.
Below, we use units with $\hbar = e = k_B = 1$.

\section{General formalism}\label{sec2}

In this section, we outline our general GF approach to the calculation of the transition rates entering the Lindblad master equation governing the dynamics of low-energy quasiparticles of a many-body quantum system.  As concrete application, we focus on the Josephson dot model introduced in Sec.~\ref{sec2a}.  Our EOM approach is then outlined in Sec.~\ref{sec2b}, followed by a calculation of all transition rates involving ABSs in Sec.~\ref{sec2c}. Finally, we specify the Lindblad equation governing the dynamics in the Andreev subspace of the Hilbert space in Sec.~\ref{sec2d}.  

\subsection{Josephson dot model}\label{sec2a}

\begin{figure}
\begin{center}
    \includegraphics[width=0.4\textwidth]{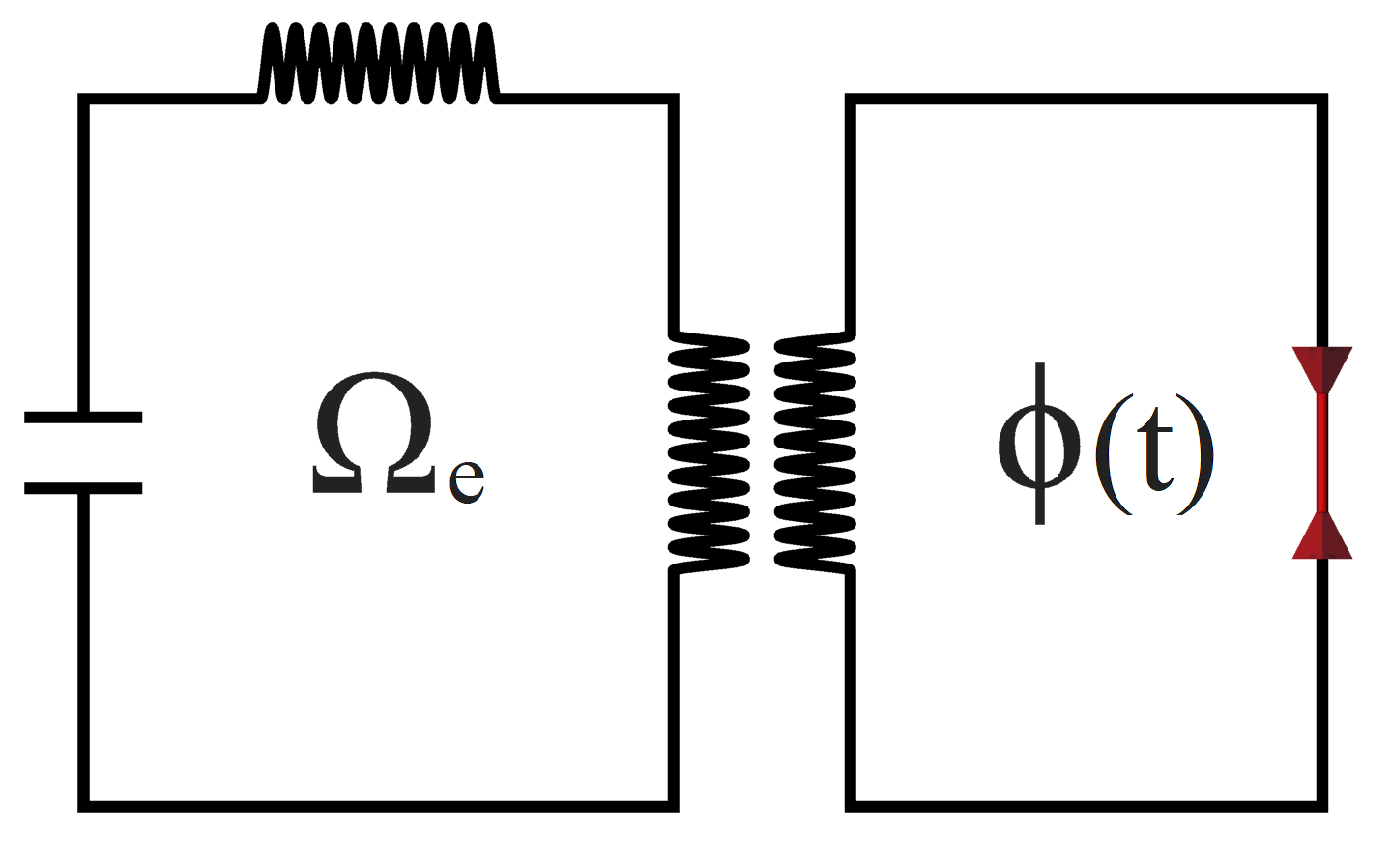}
    \caption{
     Schematic setup: 
    The average phase difference $\phi_0$ across the Josephson junction indicated by red arrows can be tuned by the magnetic flux through the right loop which is inductively coupled to 
    an LC circuit with resonance frequency $\Omega_e$. The phase difference $\phi(t)=\phi_0+\tilde\phi(t)$ across the junction thus contains a fluctuating component
    $\tilde\phi$.  For details, see main text.
    }
    \label{fig1}
\end{center}
\end{figure} 

We model the quantum dot forming the Josephson junction as a ballistic one-dimensional (1D) single-mode nanowire of length $L$ tunnel-coupled to two ($j = 1,2$) conventional $s$-wave BCS superconducting leads. 
A schematic illustration of the setup is shown in Fig.~\ref{fig1}.
The total Hamiltonian reads  $H = H_{\rm dot} + H_{\rm leads} + H_{\rm tun}$, where
the isolated dot is described by 
\begin{equation}\label{md:Hdot}
H_{\rm dot} = \int dx \, d^\dagger(x) \hat{h}(x) d(x) + H_{\rm int},\quad
d = \left( \begin{array}{c}
             d_\uparrow \\d_\downarrow
           \end{array}\right),
\end{equation}
with fermionic annihilation operators $d_\sigma(x)$ for spin projection $\sigma \in \{ \uparrow, \downarrow \}$ and 
the single-particle Hamiltonian $(\hat p=-i\partial_x)$
\begin{equation}\label{md:hath}
\hat{h}(x) =
\frac{\hat p^2}{2 m_x} - \mu + \alpha \sigma_z \hat p + {\bf b \cdot \bm \sigma} + V(x).
\end{equation}
Here $m_x$ is the effective electron mass in the wire, $\mu$ the chemical potential, the Pauli matrices $\sigma_{x,y,z}$ act in spin space, and 
the spin quantization axis is chosen along the spin-orbit direction.  The SOI coupling strength is denoted by $\alpha \geq 0$, and ${\bf b} = (b_x,0, b_z)$ is a Zeeman field including gyromagnetic and Bohr magneton factors. Due to axial symmetry, we set $b_y = 0$ without loss of generality.
Furthermore, $V(x)$ is a confinement potential along the transport direction.  In the calculations reported in 
Sec.~\ref{sec3}, we assume a hard-wall potential for $V(x)$ but one can also employ other choices.
The repulsive Coulomb interaction is described by a standard charging energy ($E_C$) term \cite{Nazarov2009},
\begin{equation}
H_{\rm int} = E_C \left( {\cal N} - n_g \right)^2,\quad
{\cal N} = \int dx \, d^\dagger(x) d(x),
\end{equation}
where $n_g$ is a dimensionless parameter proportional to a backgate voltage. The superconducting leads are described by conventional $s$-wave BCS Hamiltonians \cite{Nazarov2009} (for simplicity, we assume identical parameters for both superconductors), 
\begin{equation}\label{md:Hleads}
H_{\rm leads} = \sum_{j = 1,2} \sum_\bk \psi_{j \bk}^\dagger \left(
\xi_\bk \tau_z + \Delta \tau_x \right) \psi_{j \bk},
\end{equation}
with the Nambu spinor $\psi_{j \bk} = \left(
\begin{array}{c} \psi_{j \bk, \uparrow} \\ \psi_{j (-\bk), \downarrow}^\dagger \end{array}
\right)$. Here,
$\psi_{j \bk, \sigma}^\dagger$ creates an electron in lead $j$ with momentum $\bk$ and spin
projection $\sigma$, $\xi_\bk = k^2/(2m) - \mu$ denotes the electron dispersion in the normal-state leads, $\Delta$ is the homogeneous BCS gap, and the Pauli matrices $\tau_{x,y,z}$ act in Nambu (particle-hole) space.
Finally, with $\psi_{j \sigma} = \sum_\bk \psi_{j \bk, \sigma}$, the tunnel couplings connecting the superconducting leads to the quantum dot at the wire end points, $x=x_1$ and $x=x_2 = x_1 + L$, respectively, are described by
\begin{equation}\label{md:Htunx}
H_{\rm tun} = \sum_{j = 1,2} t_j e^{i \phi_j/2} \sum_{\sigma = \uparrow, \downarrow}
\psi_{j \sigma}^\dagger d_\sigma(x_j) + {\rm h.c.},
\end{equation}
with hopping parameters $t_j$ and superconducting phases $\phi_j$ pertaining to the respective superconductor.  For simplicity, we here assume that the tunnel couplings $t_j$ are real-valued, positive, and spin-independent parameters.
However, a complex-valued spin dependence of scattering amplitudes will arise through the SOI.  Note that we have chosen a gauge where the order parameter phase appears only in the tunneling Hamiltonian.

The noninteracting dot Hamiltonian in Eq.~\eqref{md:Hdot} can be diagonalized by a canonical transformation,
$d_\sigma(x) = \sum_\nu \chi_{\nu \sigma} (x) c_\nu$, \new{where the single-particle eigenstates $\chi_{\nu\sigma}(x)$ with eigenenergy $\epsilon_\nu$} 
describe the low-energy transport properties,  
\begin{equation}\label{doteigen}
\hat{h}(x) \chi_\nu(x) = \epsilon_\nu \chi_\nu(x),\quad
\chi_\nu = \left(
\begin{array}{c} \chi_{\nu \uparrow} \\ \chi_{\nu \downarrow} \end{array} \right).
\end{equation}
Note that \new{the quantum number} $\nu$ encapsulates both orbital and spin quantum numbers 
which cannot be disentangled in the presence of the SOI and the Zeeman field.
We apply Neumann boundary conditions, $\partial_x \chi_\nu(x_1) = \partial_x \chi_\nu(x_2) = 0$, such that no current flows through the wire ends, and choose an orthonormal basis,
$\int dx \,  \chi^\dagger_\nu(x) \chi^{}_{\nu'}(x) = \delta_{\nu\nu'}$.
\new{In practice, the Hamiltonian matrix appearing in Eq.~\eqref{doteigen} is truncated to the subspace of low-energy eigenstates below an 
energy cutoff comparable to $\Delta$, and then diagonalized numerically. For notational simplicity and convenience, we assume an even number $2 \ell$ (with integer $\ell\ge 1$) of relevant dot levels, $\nu \in \{ 1,\ldots, 2 \ell \}$, in what follows.  For the spin-degenerate case corresponding to the absence of a Zeeman field, this situation is directly realized.  Moreover, the equations below are easily adapted for an odd number of states. }

In the new basis, the dot is represented by normal-mode fermion operators, 
$c_\nu = \sum_{\sigma} \int dx \, \chi_{\nu \sigma}^\ast (x) d_\sigma(x).$
We then arrive at the $c$-fermion representation of the dot Hamiltonian,
\begin{equation}\label{md:Hdotc}
H_{\rm dot} = \sum_{\nu = 1}^{2 \ell} \epsilon_\nu c_\nu^\dagger c_\nu +
E_C \left( {\cal N} - n_g \right)^2,\quad {\cal N} = \sum_\nu c_\nu^\dagger c_\nu.
\end{equation}
Likewise, Eq.~\eqref{md:Htunx} takes the form
\begin{equation}\label{md:Htunnu}
H_{\rm tun} = \sum_{j,\sigma,\nu} e^{i \phi_j/2} t_{j \sigma, \nu} \psi_{j \sigma}^\dagger c_\nu + {\rm h.c.},\quad
t_{j \sigma, \nu} = t_j \chi_{\nu \sigma}(x_j),
\end{equation}
where the hopping parameters $t_{j \sigma, \nu}$ depend on the spin index $\sigma$ and on the 
dot level energies $\epsilon_\nu$.
Defining Nambu spinors for the boundary lead fermions $\psi_{j \sigma}$ and for the dot fermions $c_\nu$,
\begin{equation}\label{md:realconstr}
\psi_j = \left( \begin{array}{c} \psi_{j \uparrow} \\ \psi_{j \downarrow}^\dagger \end{array} \right),\quad
\gamma_\nu = \left( \begin{array}{c} c_\nu \\ c_\nu^\dagger \end{array} \right) = \tau_x \gamma_\nu^\ast,
\end{equation}
one obtains the equivalent form
\begin{eqnarray}\label{md:Htun}
H_{\rm tun} &=& \sum_{j,\nu} \psi_j^\dagger {\cal T}_{j \nu} \gamma_\nu^{} 
+ {\rm h.c.} ,\\ \nonumber
{\cal T}_{j \nu} &= &\left( \begin{array}{cc} t_{j \uparrow, \nu} e^{i \phi_j / 2} &  0 \\
0 & - t_{j \downarrow, \nu}^\ast e^{-i \phi_j / 2} \end{array} \right),
\end{eqnarray}
where the matrix structure of ${\cal T}_{j\nu}$ is in Nambu space.
Note that $\gamma_\nu$ obeys a reality constraint, see Eq.~\eqref{md:realconstr}, which implies double counting.  (In the absence of SOI and Zeeman field, double counting is not required and the formalism allows for some simplifications.  Below we consider the general case.) Correspondingly, Eq.~\eqref{md:Htun} can be written as
\begin{equation}\label{md:Htung}
H_{\rm tun} = \sum_{j,\nu} \gamma_\nu^\dagger \left( {\cal T}_{j \nu}^\ast \psi_j - \tau_x {\cal T}_{j \nu} \psi_j^\ast \right).
\end{equation}
For the application in Sec.~\ref{sec3}, we assume $E_C = 0$ (no Coulomb effects) but we discuss interacting Josephson dots using this formalism elsewhere. In Nambu notation, up to an irrelevant constant term, the dot Hamiltonian \eqref{md:Hdotc} is then given by
\begin{equation}\label{md:Hdotg}
H_{\rm dot} = \frac{1}{2}\sum_{\nu = 1}^{2 \ell} \epsilon_\nu  \gamma_\nu^\dagger \tau_z \gamma_\nu^{}.
\end{equation}

We assume below that the nanowire is embedded in a superconducting loop threaded by a magnetic flux, 
see Fig.~\ref{fig1}. The superconducting phase difference across the nanowire, $\phi = \phi_1 - \phi_2$,
is treated as dynamical variable, $\phi = \phi_0 + \tilde{\phi},$ 
where $\phi_0$ is the average phase difference induced by the magnetic flux and $\tilde{\phi}$ is a bosonic operator. 
This operator is time-independent in the Schr\"odinger picture and represents the fluctuating phase caused by the electromagnetic environment, 
e.g., a circuit resistance or the inductive coupling to a microwave resonator.
Without loss of generality, we write
\begin{equation} \label{sdef}
  \phi_j = s_j \phi/2,\quad s_1 = 2 + s_2 \in (0, 2).
\end{equation}
Note that we work in a gauge with vanishing vector potential in the nanowire region, see Eq.~\eqref{md:hath}. 
In addition, for nanowires of length $L \gtrsim \xi_0$, where $\xi_0 = v_{\rm F} / \Delta$ is the superconducting coherence length ($v_{\rm F}$ is the Fermi velocity in the leads), Josephson transport is in general
sensitive to the phase shift asymmetry $s_1/s_2$. In the symmetric case, one has $s_1 = -s_2 = 1$.  An asymmetry 
of the phase shift can be associated with the capacitive asymmetry of the tunnel contacts at $x_{1,2}$. 

Modeling the electromagnetic environment by a continuum of harmonic oscillators (boson modes) in thermal equilibrium at temperature $T_b$,
in the Heisenberg picture, phase fluctuations $\tilde{\phi}(t)$ are characterized by the bath correlation function
\new{
\begin{equation}\label{md:calD}
  {\cal D}(t,t') = \left\langle \frac{\tilde\phi(t)}{2} \frac{\tilde\phi(t')}{2} \right\rangle_b = \int_{-\infty}^\infty\frac{d \omega}{2 \pi} \, e^{-i \omega (t - t')} {\cal D}(\omega).
\end{equation}
With $\langle \tilde{\phi}(t) \rangle_b = 0$, where $\langle \cdots \rangle_b$ denotes 
a thermal average using the Bose-Einstein function
$n_B(\omega) = 1/(e^{\omega / T_b} - 1)$, one can write \cite{Weiss_2012}
\begin{equation}\label{md:calDom}
{\cal D}(\omega) = 2 \pi J(\omega) \left[ n_B(\omega) + 1 \right],
\end{equation}
where $J(\omega) = -J(-\omega)$ denotes the \emph{spectral density} of the environment, 
with $J(\omega > 0) \geq 0$.   We give examples for $J(\omega)$} in the setup of Fig.~\ref{fig1} later on, see Eqs.~\eqref{jw} and \eqref{jw_ohm}. 

\subsection{EOM approach}\label{sec2b}

We next formulate an EOM approach to the dynamics of the system described in Sec.~\ref{sec2a}. 
The Heisenberg EOMs for the Nambu fermion field operators $\psi_{j\bk}(t)$ and $\gamma_\nu(t)$ are given by, see Eqs.~\eqref{md:Hleads} and \eqref{md:Htung}--\eqref{md:Hdotg},
\begin{equation}\label{eom:eompsijk}
  \left( i \partial_t - h_\bk \right) \psi_{j\bk}(t) = \sum_{\nu=1}^{2\ell} {\cal T}_{j \nu}(t) \gamma_\nu(t),
\end{equation}
with $h_\bk = \xi_\bk \tau_z + \Delta \tau_x$  and
\begin{equation}\label{eom:eomgmnu}
  \left( i \partial_t - \epsilon_\nu \tau_z \right) \gamma_\nu(t) =
  \sum_{j=1,2} \left[ {\cal T}_{j \nu}^\ast(t) \psi_j(t) - \tau_x {\cal T}_{j \nu}(t) \psi_j^\ast(t) \right]  ,
\end{equation}
with $\psi_j(t) = \sum_\bk \psi_{j \bk}(t)$.  Here, ${\cal T}_{j \nu}(t)$ follows from the corresponding expression in Eq.~\eqref{md:Htun} by letting $\phi_j\to \phi_j(t)=s_j\phi(t)/2$.
For time-independent average phase bias $\phi_0$, we have $\phi(t)=\phi_0+\tilde{\phi}(t)$ in the Heisenberg picture.

From Eq.~\eqref{eom:eompsijk}, we infer that the retarded response of the boundary fermions $\psi_j(t)$ to the dot dynamics and to the phase fluctuations is given by
\begin{equation}\label{eom:psij}
  \psi_j(t) = \int dt' \, g^R(t,t') \sum_\nu {\cal T}_{j \nu}(t') \gamma_\nu(t'),
\end{equation}
where $g^R(t,t')$ is the retarded boundary GF of the leads,
\begin{eqnarray} \label{gfleads}
  g^R(t,t') &=& \int \frac{d \omega}{2\pi} \, e^{-i \omega (t-t')} g^R(\omega),\\ \nonumber
  g^R(\omega) &=& \sum_\bk \frac{1}{\omega - h_\bk + i0^+} = - \pi \nu_F
  \frac{ \omega \tau_0 + \Delta \tau_x}{\zeta(\omega)},
\end{eqnarray}
with the Nambu-space identity matrix $\tau_0$, the normal-state ($\Delta=0)$ lead density of states $\nu_F = \sum_\bk \delta(\xi_\bk)$, and the auxiliary function
\begin{eqnarray}\label{zeta}
  \zeta(\omega) &=& \sqrt{\Delta^2 - (\omega + i 0^+)^2} \\ &=&
\left\{ \begin{array}{cc} \sqrt{\Delta^2 - \omega^2}, & |\omega| \le \Delta, \\ \nonumber
 - i \, {\rm sgn}(\omega) \sqrt{\omega^2 - \Delta^2}, & |\omega| > \Delta. \end{array} \right.
\end{eqnarray}
We often keep $\tau_0$ implicit in what follows.

We  next insert Eq.~\eqref{eom:psij} into Eq.~\eqref{eom:eomgmnu}. As a result, 
with a time-nonlocal matrix kernel acting in level-Nambu space,
\begin{eqnarray} \nonumber
  \tilde \Lambda_{\nu \nu'}(t,t') &=& \sum_{j=1,2} \left( \tilde W_{j, \nu \nu'}(t,t') -
  \tau_x \tilde W_{j, \nu \nu'}^\ast(t,t') \tau_x \right),\\ \label{Wjdef}
  \tilde W_{j, \nu \nu'}(t,t') &=& {\cal T}_{j \nu}^\ast(t) g^R(t,t') {\cal T}_{j \nu'}(t'),
\end{eqnarray}
we arrive at a closed set of EOMs for the dot fermions,
\begin{equation}\label{eom:eom}
  \left( i \partial_t - \epsilon_\nu \tau_z \right) \gamma_\nu(t) =
  \sum_{\nu'=1}^{2 \ell} \int dt' \tilde \Lambda_{\nu \nu'}(t,t') \gamma_{\nu'}(t'),
\end{equation}
which are nonlocal in time.  The above steps are equivalent to integrating out the lead fermions on the retarded branch of the Keldysh contour \cite{Weiss_2012}. The tilde notation emphasizes that $\tilde{W}_j$ and $\tilde \Lambda$ depend on the phase fluctuations $\tilde{\phi}$ due to the boson modes. 

Below we assume the weak-coupling limit for the electromagnetic environment. 
Treating $\tilde{\phi}$ as  small perturbation, ${\langle \tilde{\phi}^2 \rangle}_b \ll 1$, we expand ${\cal T}_{j \nu}(t)$ to first order in $\tilde{\phi}(t)$ using, see Eqs.~\eqref{md:Htun} and \eqref{sdef},
\begin{equation}
   e^{ \pm i\phi_j(t)/2}= e^{\pm is_j \phi_0 / 4} \left(
  1 \pm \frac{i}{4}  s_j \tilde{\phi}(t) + o(\tilde{\phi})
  \right).
\end{equation}
It is now convenient to introduce a $4\ell$-component multispinor field $\gamma=\left( \gamma_1, \ldots, \gamma_{2\ell} \right)^T$ combining all Nambu bispinors $\gamma_\nu$, such that Eq.~\eqref{eom:eom} takes the form 
\begin{equation}\label{eom:eomgm}
  \left( i \partial_t - \epsilon \tau_z \right) \gamma(t) =
  \int dt' \tilde \Lambda(t,t') \gamma(t'), 
\end{equation}
where $\tilde \Lambda(t,t')$ and $\tilde{W}_j(t,t')$ are $4 \ell \times 4 \ell$ matrices in level-Nambu space, see
Eq.~\eqref{Wjdef}.
We also define the $2\ell$-dimensional diagonal matrix
$\epsilon = {\rm diag}\left(\epsilon_1, \ldots, \epsilon_{2 \ell}\right)$, and 
$2 \ell \times 2\ell$ hybridization matrices ($j=1,2)$ in level space,
\begin{eqnarray}
\label{hybrid}
    \Gamma_{j, \nu' \nu} &=&
    \pi \nu_F \sum_\sigma t_{j \sigma, \nu'}^\ast t_{j \sigma, \nu},\\ \nonumber
    F_{j, \nu' \nu} &=& \pi \nu_F
    \sum_\sigma \sigma t_{j \sigma, \nu'} t_{j (-\sigma), \nu},
\end{eqnarray}
with $t_{j\sigma,\nu}$ in Eq.~\eqref{md:Htunnu} and $\sigma = \uparrow /\downarrow=+/-$.  

\new{We note} that the matrices $\Gamma_j$ are Hermitian,
  $\Gamma_j = \Gamma_j^\dagger$, while the $F_j$ are antisymmetric, $F_j = -F _j^T$.
In the appendix, we provide technical details on the calculation of the wave functions $\chi_{\nu}(x)$ for the present model. These wave function are used in 
Sec.~\ref{sec3}, in particular, to compute the hybridization matrices in Eq.~\eqref{hybrid}.
  
After some algebra,  to leading order in $\tilde{\phi}(t)$, Eq.~\eqref{eom:eomgm} can be written as 
\begin{equation}\label{eom:eomPT}
  \int dt' {\cal L}(t,t') \gamma(t') = \int dt' \tilde V(t,t') \gamma(t'),
\end{equation}
where the Lagrangian kernel ${\cal L}(t,t')$ describes the noninteracting dynamics of the junction, both for ABSs and above-gap continuum excitations, in the absence of phase fluctuations,
\begin{equation}
  {\cal L}(t,t') = \int \frac{d \omega}{2\pi} \,
  e^{-i\omega(t-t')} {\cal L}(\omega),\quad
  {\cal L}(\omega) = \omega - \epsilon \tau_z - \Lambda(\omega),
\end{equation}
with $\Lambda(\omega) = \sum_{j} \Lambda_j(\omega)$. 
Using Eq.~\eqref{zeta}, we obtain the Nambu matrix structure (the dot level structure is encoded in $\epsilon$, $\Gamma_j$ and
$F_j$)
\begin{equation}\label{eom:Lmbd}
  \Lambda_j(\omega) = -\frac{1}{\zeta(\omega)}\left( \begin{array}{cc}
                        \omega \Gamma_j & \Delta e^{- i s_j \phi_0/2} F_j^\dagger \\
                        \Delta e^{i s_j \phi_0/2} F_j & \omega \Gamma_j^\ast
                      \end{array} \right).
\end{equation}
The kernel $\tilde V(t,t')$ in Eq.~\eqref{eom:eomPT} describes the linear-in-$\tilde{\phi}$ coupling to the electromagnetic environment,
\begin{equation}\label{eom:tldV}
  \tilde V(t,t') = \frac{\tilde{\phi}(t)}{2} \tau_z {\cal I}(t,t') - {\cal I}(t,t') \tau_z \frac{\tilde{\phi}(t')}{2},
\end{equation}
with the time-nonlocal current operator
\begin{equation}\label{eom:calI}
  {\cal I}(t,t') = \int \frac{d \omega}{2\pi} \,
  e^{-i\omega(t-t')} {\cal I}(\omega),\quad
  {\cal I}(\omega) = \frac{1}{2i}\sum_{j} s_j \Lambda_j(\omega).
\end{equation}

\subsubsection{No fluctuations}

In the absence of phase fluctuations, $\tilde{\phi} = 0$,  the above equations simplify to
\begin{equation}\label{eom:eomh}
    \left[ i \partial_t - h(i \partial_t) \right] \gamma(t) = 0,\quad
    h(\omega) = \epsilon \tau_z + \Lambda(\omega),
\end{equation}
where $h(i \partial_t)$ plays the role of an effective single-particle Hamiltonian. However,
this operator is nonlocal in time since it can be expanded into an infinite series in $\partial_t$. 
In particular, we observe from Eqs.~\eqref{zeta} and \eqref{eom:Lmbd} that $h(\omega)$ is Hermitian, $h(\omega) = h^\dagger(\omega)$,  only for energies within the subgap region $|\omega|<\Delta$.

In the \emph{atomic limit}, which is described by a very large pairing gap (formally, $\Delta \rightarrow \infty$, such that continuum states can be disregarded \cite{Alvaro2011}),
one obtains the frequency-independent Hermitian Hamiltonian  
\begin{equation}\label{eom:hA}
     h_A =     \epsilon \tau_z - \sum_{j=1,2}
     \left( \begin{array}{cc} 0 & F_j^\dagger e^{-i s_j \phi_0/2} \\
     F_j e^{i s_j \phi_0/2} & 0 \end{array} \right).
\end{equation}
Clearly, we then have only ABS solutions. 

For finite $\Delta$, the solution of Eq.~\eqref{eom:eomh} can be written as a series expansion in terms of quasiparticle field operators involving either ABSs or above-gap continuum states,
\begin{eqnarray} \nonumber
  \gamma(t) &=& \tau_x \gamma^\ast(t) = \sum_{\nu = 1}^{2 \ell} \left(
  a_\nu \eta_\nu e^{- i E_\nu t} + a_\nu^\dagger \tau_x \eta_\nu^\ast e^{i E_\nu t}
  \right) \\ 
  &+& \int_{|\omega| > \Delta} \frac{d \omega}{2 \pi} \, e^{- i \omega t}
  \tilde \gamma_\omega,
  \quad \tilde \gamma_{-\omega} = \tau_x \tilde \gamma_\omega^\ast,\label{eom:gmexpand}
\end{eqnarray}
where the $a_\nu$ are fermion annihilation operators for ABSs with energy $E_\nu$ with $|E_\nu|<\Delta$ and eigenspinor $\eta_\nu$ in level-Nambu space,
\begin{equation}\label{eom:ALScrheq}
  h(E_\nu) \eta_\nu = E_\nu \eta_\nu.
\end{equation}
As to the number of subgap solutions, or localized states in a more general context, we assume that it is $2\ell$ (without double counting) based on the continuity of the ``root flow'' in $\Gamma$-$\Delta$ parameter space. 
In fact, the number of roots is $2\ell$ in two limits, namely $\Gamma=0$ or $\Delta\to \infty$. We expect
that this number remains $2\ell$ when increasing (decreasing) $\Gamma$ ($\Delta$). 
The only exception is the point $\Delta=0$, where a phase transition occurs. As a consequence, the dot level energies $\epsilon_\nu$ do not need to lie below $\Delta$. For high-energy dot states, the corresponding Andreev levels are expected to merge with the BCS gap edges.

The field operator $\tilde \gamma_\omega$ represents continuum states at energy $\omega$, where a particle-hole symmetry relation is imposed by double counting,
\begin{equation}\label{eom:PHsymm}
  h(\omega) = - \tau_x h^\ast(-\omega) \tau_x.
\end{equation}
A few remarks are now in order. First, in contrast to the ABS modes $\propto e^{\pm i E_\nu t}$, the continuum harmonics $\tilde \gamma_\omega e^{- i \omega t}$ cannot be defined as eigenstate solutions of Eq.~\eqref{eom:eomh}, since $h(\omega)$ is non-Hermitian for $|\omega| > \Delta$.  
Second, the Andreev eigenspinors $\eta_\nu$ can always be normalized, $\eta^\dagger_\nu \eta_\nu = 1$. However, in general, they are not necessarily orthogonal for different ABSs,
i.e., ${\eta}^\dagger_\nu \eta_{\nu'} \neq 0$ for $\nu \neq \nu'$, since $[h(E_\nu), h(E_{\nu'})] \neq 0$ except in the atomic limit.

\subsubsection{Phase fluctuation effects}

We now return to the full EOM \eqref{eom:eomPT} in the presence of phase fluctuations $\tilde{\phi}$.
Since Eq.~\eqref{eom:eomPT} is linear in the field operator $\gamma(t)$, it is convenient to switch to a first-quantized framework for the fermionic part. In first quantization, ABS and continuum quasiparticles are represented by a $4\ell$-component wave function $\Psi(t)$, which obeys a time-nonlocal Schr\"odinger equation,
\begin{equation}\label{eom:Schreq}
    {\cal L}(i \partial_t) \Psi(t) = \int dt' \tilde V(t,t') \Psi(t'),\quad
    {\cal L}(i \partial_t) = i \partial_t - h(i \partial_t).
\end{equation}
The perturbation $\tilde{V}(t,t')$ is linear in the bosonic operator $\tilde{\phi}(t)$, see Eq.~\eqref{eom:tldV}. The latter plays the role of an external fluctuating force and has the
correlation function ${\cal D}(t,t')$ in Eq.~\eqref{md:calD}. 
We note that $\left[ \tilde{\phi}(t), \tilde{\phi}(t') \right] \neq 0$ for $t \neq t'$, and hence \new{${\cal D}(\omega) \neq {\cal D}(-\omega)$,} cf. Eq.~\eqref{md:calDom}, except in the classical oscillator limit corresponding to high temperature \new{$T_b\gg \omega$}. However, beyond this limit, any solution $\Psi(t)$ of Eq.~\eqref{eom:Schreq} is still an operator with respect to the bosonic bath.

Similar to Eq.~\eqref{eom:gmexpand}, the expansion of $\Psi(t)$ in the quasiparticle basis can be written as
\new{
\begin{eqnarray}\nonumber
  \Psi(t) = \tau_x \Psi^\ast(t) &=& \sum_{\nu = 1}^{2 \ell} \left[
  a_\nu(t) \eta_\nu e^{- i E_\nu t} + a_\nu^\ast(t) \tau_x \eta_\nu^\ast e^{i E_\nu t}
  \right]\\  &+& \tilde{\Psi}(t),\label{eom:Psiexpand}
\end{eqnarray}
}where in the first-quantized framework, the $a_\nu(t)$ are now complex-valued  probability amplitudes for ABSs, see~Eq.~\eqref{eom:ALScrheq}. Their time dependence comes from $\tilde{\phi}(t)$. 
Likewise, $\tilde{\Psi}(t)$ represents the continuum harmonics with $|\omega| > \Delta$.
In what follows, without loss of generality, we assume $E_\nu > 0$ for all $\nu \in \{1,\ldots, 2\ell \}$, where double-counting partners with negative energy (related by particle-hole symmetry) are labeled by $\bar{\nu}$, i.e., $E_{\bar \nu} = - E_\nu$ and $\eta_{\bar \nu} = \tau_x \eta_\nu^\ast$.  We emphasize that
no redefinition of $\ell$ is required. With double counting, we thus have $4\ell$ states.
For simplicity, we here assume that no zero modes with $E_\nu=0$ are present.  The corresponding modifications
necessary to describe such situations are straightforward to implement.

To compute transition rates for quasiparticle states in the presence of $\tilde{\phi}(t)$, our strategy is to solve Eq.~\eqref{eom:Schreq} iteratively to first order in $\tilde{V}$. We then calculate the transition probability to a given state by averaging over the phase fluctuations, assuming that the bath remains in thermal equilibrium at temperature $T_b$ at all times. Besides, we assume that ABSs (with energies 
$\pm E_\nu$) are not entangled with continuum states ($|\omega|>\Delta$) \cite{Zazunov2014}, i.e., continuum states 
play the role of a fermionic bath for the ABS sector.
We model the continuum state distribution by a thermal quasi-equilibrium Fermi function with an effective ``quasiparticle temperature'' $T_{\rm qp}$. This temperature is set by the temperature of the BCS superconducting leads, and can in general be different from the environmental temperature $T_b$.  
In Sec.~\ref{sec3}, we consider the regime $T_{\rm qp}\ge T_b$.

We note that an extension of our approach to include thermal phonons in the effective EOM \eqref{eom:eomPT} is quite straightforward. It can be achieved by adding the corresponding interaction term $H_{\rm e-ph}$ to the full Hamiltonian $H$. For instance, $H_{\rm e-ph}$ can describe the lead electrons $\psi_j(\br)$ coupled to longitudinal acoustic phonons within the deformation potential approximation, $H_{\rm e-ph} \sim \int d\br \, \psi_j^\dagger(\br) \tau_z \psi_j(\br) \nabla \cdot {\bf u}(\br)$, where ${\bf u}(\br)$ is the phonon displacement field operator. Following the steps in Sec.~\ref{sec2b} then leads to Eq.~\eqref{eom:eomPT}, with the right-hand-side containing along with the phase-fluctuation kernel $\tilde{V} \propto \tilde{\phi}$ also a similar (non-local in time) term 
$V_{\rm ph} \propto \xi$. This term is  linear in a dimensionless bosonic variable $\xi$ encapsulating the phonon modes. The fluctuations $\tilde{\phi}$ vs $\xi$ play the role of an external vs intrinsic bosonic environment for the dot fermions.  These environments equilibrate at the temperatures $T_b$ and $T_{\rm qp}$, respectively. 
As shown in Sec.~\ref{sec3}, the condition $T_{\rm qp} > T_b$ is an important ingredient for observing QMEs in our setup. 
Although the original coupling to the $\tilde{\phi}$ fluctuations is mainly restricted to the dot region, phonon modes interact with lead electrons throughout the junction. (For short nanowires, 
the direct electron-phonon interaction in the dot region can be neglected due to the small size of the corresponding spatial region.) The chosen temperature regime ($T_{\rm qp} > T_b$) implies local cooling of the dot region, which can be realized if the coupling of the dot to phase fluctuations is stronger than its coupling to phonons.  As a consequence, in the nanowire dot region,
local equilibration to a thermal state at temperature $\approx T_b$ is then expected.

For superconducting leads, the relaxation rate (inverse lifetime) $\tau_{\rm ph}^{-1}$ of bulk quasiparticles due to electron-phonon interactions can be estimated as
the relaxation rate of a normal-state electron with energy $\Delta$ above the Fermi level. One finds  $\tau_{\rm ph}^{-1} \sim \Delta^3 / \Theta_D^2$ \cite{Kaplan1976}, where $\Theta_D$ is the Debye temperature. This estimate gives an upper bound on the phonon-induced relaxation rate for ABSs. In short junctions, this rate is even smaller by a (generally geometry-dependent) factor $L/\xi_0$, as shown in Ref.~\cite{alqB} for an adiabatic constriction model. The relaxation rate due to phase fluctuations can be estimated as $\tau_\phi^{-1} \sim \Delta^2 \omega_\phi$, where $\omega_\phi \sim J(\Delta)$ is the characteristic energy scale of the spectral density $J(\omega)$ at relevant transition energies. 
In particular, for the Lorentzian shape \eqref{jw}, we estimate $\omega_\phi\sim \eta / \Delta^2$ in the regime $\Omega_e, \eta \ll \Delta$, where $\eta$ is the damping strength.
As a result, electromagnetic phase fluctuations strongly dominate over phonon relaxation processes, $\tau_\phi^{-1} \gg \tau_{\rm ph}^{-1}$, for sufficiently strong damping of the LC oscillator,
\begin{equation}
\eta / \Delta \gg (L/\xi_0) (\Delta / \Theta_D)^2.
\end{equation}
For instance, for a short junction with Al leads and  $L/\xi_0 \sim 10^{-1}$,  taking the Lorentzian spectral density \eqref{jw} of an LC circuit, we  estimate
$\eta \gg 10^{-5} \Delta$. Assuming that this condition is met, phonon-induced processes are henceforth considered as subleading and will not be taken into account.

Let us then consider the time evolution of the Andreev state $\eta_\lambda$ in the presence of phase fluctuations. To first order in $\tilde{V}$, the solution of the Schr\"odinger equation \eqref{eom:Schreq} is 
given by $\Psi(t) = \eta_\lambda e^{- i E_\lambda t} + \Psi^{(1)}_\lambda(t)$ with
\begin{equation}\label{eom:Psit}
  \Psi^{(1)}_\lambda(t) = \int dt_1 dt_2 \, G^R(t,t_1) \tilde{V}(t_1, t_2) \eta_\lambda e^{- i E_\lambda t_2},
\end{equation}
where $G^R(t,t')$ is the retarded GF of the junction in the absence of phase fluctuations,
\begin{eqnarray}\nonumber  
    G^R(t,t') &=& \int \frac{d\omega}{2\pi} \, e^{-i \omega (t-t')} G^R(\omega),\\
    G^R(\omega) &=& \frac{1}{\omega - h(\omega) + i \delta_+}.\label{eom:GR}
\end{eqnarray}
We put $\delta_+ \rightarrow 0^+$ at the end of the calculation but allow for a finite phenomenological
quasiparticle decay rate $\delta_+$ in intermediate steps. Using Eq.~\eqref{eom:tldV} with
\new{
$\tilde{\phi}(t) = \int \frac{d\omega}{2\pi} \, e^{-i \omega t} \phi_\omega$, where 
$\phi_\omega^\dagger = \phi_{-\omega}^{}$, we obtain
\begin{eqnarray}\nonumber 
  \Psi^{(1)}_\lambda(t) &=& \int \frac{d\omega}{2\pi} \, e^{-i(E_\lambda + \omega)t} G^R(E_\lambda + \omega) \\ &\times& \left[
  \tau_z {\cal I}(E_\lambda) - {\cal I}(E_\lambda + \omega) \tau_z
  \right] \eta_\lambda \frac{\phi_\omega}{2}. \label{eom:Psi1}
\end{eqnarray}
}In Sec.~\ref{sec2c}, we use this result in order to compute transition rates involving ABSs.  Those transition rates in turn appear in the Lindblad equation governing the dynamics of the ABS sector, see Sec.~\ref{sec2d}.

\subsection{Transition rates}\label{sec2c} 

In this subsection, we discuss the transition rates involving ABSs which are induced by the electromagnetic environment.  We compute rates connecting different ABSs as well as those between ABS and above-gap continuum states.

\subsubsection{Atomic limit}

Before tackling the full expression \eqref{eom:Psi1}, it is instructive to first study the atomic limit, where 
substantial simplifications are possible. Taking $\Delta \rightarrow \infty$, see Eq.~\eqref{eom:hA}, we find
\begin{equation}\label{al:GR}
    G^R(\omega)= \sum_{\nu = 1}^{2\ell}
    \left[\frac{\eta_\nu \eta^\dagger_\nu}{\omega - E_\nu + i \delta_+} +
    \frac{\eta^{}_{\bar\nu}{\eta}^\dagger_{\bar \nu}}{\omega + E_\nu + i \delta_+}
    \right],
\end{equation}
where we used the completeness of the ABS {\em orthonormal} basis. In this limit, the perturbation term \eqref{eom:tldV} reduces to
$\tilde V(t,t') = \frac{\tilde{\phi}(t)}{2} I_A \delta(t-t'),$
with the time-independent supercurrent operator $I_A = \left[ \tau_z, {\cal I} \right]$.  Using Eq.~\eqref{eom:calI}, we find
\begin{equation}
  I_A = \sum_{j} s_j
    \left( \begin{array}{cc} 0 & i F_j^\dagger e^{-i s_j \phi_0/2} \\
     -i F_j e^{i s_j \phi_0/2} & 0 \end{array} \right) = 2 \frac{\partial h_A}{\partial \phi_0}.
\end{equation}
In general, $\left[ I_A, h_A \right] \neq 0$. One exception is the ballistic limit of perfect transparency, which corresponds to $\epsilon = 0$, symmetric hopping amplitudes $t_1=t_2$, and 
$s_1 = -s_2 = 1$ in Eq.~\eqref{sdef}. In this case, with $F_1 = F_2 = F/2$, one finds
\begin{equation} 
  h_A = - \left( \begin{array}{cc} 0 & F^\dagger \\
     F & 0 \end{array} \right) \cos\frac{\phi_0}{2},\quad
  I_A = \left( \begin{array}{cc} 0 & F^\dagger \\
     F & 0 \end{array} \right) \sin\frac{\phi_0}{2}     ,
\end{equation}
and therefore $\left[ I_A, h_A \right] = 0$. Phase fluctuations thus cannot induce transitions between ballistic ABSs.

Applying the atomic limit to Eq.~\eqref{eom:Psi1} for general junction parameters, we obtain
\new{
\begin{eqnarray}\nonumber
  \Psi^{(1)}_\lambda(t) &=& \int \frac{d\omega}{2\pi} \, e^{-i(E_\lambda + \omega)t}
  G^R(E_\lambda + \omega) I_A \eta_\lambda \frac{\phi_\omega}{2} \\
  &=& \sum_{\nu = 1}^{2 \ell} \left[
  a_\nu(t) \eta_\nu e^{- i E_\nu t} + a_{\bar \nu}(t) \eta_{\bar \nu} e^{i E_{\bar \nu} t}
  \right].
  \label{eom:Psi1expand}
\end{eqnarray} }
Note that here $a_\nu(t)$ and $a_{\bar \nu}(t)$ are independent probability amplitudes, in contrast to Eq.~\eqref{eom:Psiexpand}, where $a_{\bar \nu}(t) = a_\nu^\ast(t)$ due to the imposed reality constraint $\Psi(t) = \tau_x \Psi^\ast(t)$. The point is that in Eq.~\eqref{eom:Psit} we consider a scattering problem with an incoming state $\Psi_{\rm in}(t) = \eta_\lambda e^{- i E_\lambda t}$ which breaks  particle-hole symmetry, $\Psi_{\rm in}(t) \neq \tau_x \Psi_{\rm in}^\ast(t)$. Accordingly, the outgoing solution $\Psi^{(1)}_\lambda(t)$ is not required to obey the reality constraint.
In Eq.~\eqref{eom:Psi1expand}, the amplitude $a_n(t)$ with $n \in \{ \nu, \bar{\nu} \}$  is given by
$a_n(t) = e^{i E_n t} \eta^\dagger_n \Psi^{(1)}_\lambda(t)$.  Consequently, the probability for an inter-level transition $\lambda \rightarrow n \neq \lambda$, averaged over phase fluctuations, 
is given by
\new{
\begin{eqnarray}\label{al:wlmbdn}
&&  w_{\lambda \rightarrow n}(t)  =  \langle  a_n^\ast(t)  a_n(t) \rangle_b \\ \nonumber &&=
  \int \frac{d \omega d \omega'}{(2\pi)^2} \, e^{-i (\omega - \omega') t}
  \left\langle \frac{\phi_{\omega'}^\dagger}{2} \frac{\phi_\omega}{2} \right\rangle_b \\ \nonumber
  &&\times \left[ \eta^\dagger_\lambda I_A G^A(E_\lambda + \omega') \eta_n \right]
  \left[ \eta^\dagger_n G^R(E_\lambda + \omega) I_A \eta_\lambda \right],
\end{eqnarray}
where $G^A(E) = \left[ G^R(E) \right]^\dagger$} is the advanced GF. Using Eqs.~\eqref{md:calD}, \eqref{md:calDom},
and \eqref{al:GR}, we  obtain 
\new{
\begin{equation}\label{al:Ilmbdn}
  w_{\lambda \rightarrow n} = \int d\omega \, J(\omega) n_B(\omega)
  \frac{|I_{n \lambda}|^2}{\left| E_\lambda + \omega - E_n + i \delta_+ \right|^2},
\end{equation}
}with $I_{n \lambda} =  \eta^\dagger_n I_A \eta_\lambda$.
For $\delta_+ \rightarrow 0^+$, we have
$\frac{1}{| x + i \delta_+ |^2}  \to 2 \pi \tau_+ \delta(x)$ with the time scale
$\tau_+ = \frac{1}{2 \delta_+}\to \infty$.  This time scale corresponds to  the
 decay time of the transition probability, $w_{\lambda \rightarrow n}(t) \propto e^{- t/\tau_+}$, as follows from $G^R(t,0) \propto \Theta(t) e^{- \delta_+ t}$ before averaging over the bath (where $\Theta$ is the Heaviside step function).
For infinitely long observation time, we expect $\tau_+ \rightarrow \infty$ if no intrinsic sources of dissipation are present in the junction ($\delta_+ \rightarrow 0^+$). Hence the transition probability per unit time, i.e., the transition rate, is given by
\new{
\begin{equation}\label{al:Glmbdn}
  \Gamma_{\lambda \to n}   = \frac{w_{\lambda \to n}}{\tau_+} =
  2\pi |I_{n \lambda}|^2 J(\omega) n_B(\omega) \Big|_{\omega = E_n - E_\lambda}.
\end{equation}
}Equation~\eqref{al:Glmbdn} reproduces the Fermi golden rule result obtained in the atomic limit along the BdG route, e.g., in Refs.~\cite{Zazunov2014,Ackermann2023,Zatsarynna2024}. Here, this result 
has instead been derived from the GF approach by solving the Schr\"odinger equation \eqref{eom:Schreq}, without  explicit construction of BdG eigenstates. 
The cases $E_n - E_\lambda > 0$ and $E_n - E_\lambda < 0$ describe transitions $\lambda \rightarrow n$ induced by the absorption and emission of a boson, respectively.

We next extend the calculation of transition rates beyond the atomic limit. We recall that
for finite $\Delta$, the Schr\"odinger equation  \eqref{eom:Schreq} is nonlocal in time. 
We first compute all transition rates within the ABS sector, and then
those connecting ABSs and continuum states.

\subsubsection{Transition rates between Andreev states}

We return to the scattering problem \eqref{eom:Psit} with $\Psi^{(1)}_\lambda(t)$ in Eq.~\eqref{eom:Psi1}, where we may write
\begin{equation}\label{tr:Psi1}
  \Psi^{(1)}_\lambda(t)  =  
  \sum_{n \in \{ \nu, \bar{\nu} \}} a_n(t) \eta_n e^{- i E_n t} + \tilde{\Psi}(t),\quad
  \eta_{\bar \nu} = \tau_x \eta_\nu^\ast,
\end{equation}
with $\nu \in \{ 1, \ldots, 2\ell\}$. We recall that $\tilde{\Psi}(t)$ represents above-gap continuum states and that, in general, ${\eta}^\dagger_n \eta_{n'} \neq 0$ for $n \neq n'$.  For very long times $t=T$, $a_n(t)$ approaches the constant value $a_n =  \frac{1}{T}\int_{0}^{T} dt\, e^{i E_n t} {\eta}^\dagger_n \Psi^{(1)}_\lambda(t)$, resulting in
\new{
\begin{eqnarray}\label{as:anT}
  a_n &=&  \int \frac{d\omega}{2\pi} \, \frac{\phi_\omega}{2} \, Q_T^\ast(E_\lambda + \omega - E_n) 
     \\ \nonumber &\times& \eta^\dagger_n G^R(E_\lambda+\omega) \, \left[
  \tau_z {\cal I}(E_\lambda) - {\cal I}(E_\lambda + \omega) \tau_z
  \right] \eta_\lambda,
\end{eqnarray}
with the quantity
\begin{equation}\label{as:QT}
  Q_T(E) \equiv \frac{1}{T}\int_{0}^{T} dt\, e^{i E t} =
  \frac{e^{i E T} - 1}{i E T} .
\end{equation}
This function, for finite but large time $T$, can be regarded as a continuous version of the Kronecker symbol $\delta_{E, 0}$, i.e., it is not a singular function of $E$. 
Note that $Q_T(E=0)=1$ but $Q_T(E\ne 0)\to 0$ for $T\to \infty$.}

For \new{$E_\lambda + \omega \simeq E_n$,} which is realized to good accuracy because of the $Q_T$-factor in Eq.~\eqref{as:anT}, one finds, see Eqs.~\eqref{eom:ALScrheq} and \eqref{eom:GR},
\new{
\begin{eqnarray}\nonumber
  \eta^\dagger_n G^R(E_\lambda + \omega) &\simeq& \eta^\dagger_n \frac{1}{E_\lambda + \omega - h(E_n) + i \delta_+} \\
  &=&
  \frac{1}{E_\lambda + \omega - E_n + i \delta_+} \eta^\dagger_n.
\end{eqnarray}
As a result, we find
\begin{equation}
  a_n = \int \frac{d\omega}{2\pi} \, \frac{\phi_\omega}{2} \, Q_T^\ast(E_\lambda + \omega - E_n) \,
  \frac{I_{n \lambda}}{E_\lambda + \omega - E_n + i \delta_+},
\end{equation}
}with the current matrix element
\begin{equation}\label{as:Ilmbdn}
  I_{n \lambda} = \eta^\dagger_n \left[
  \tau_z {\cal I}(E_\lambda) - {\cal I}(E_n) \tau_z \right] \eta_\lambda.
\end{equation}

Proceeding now along the same steps as in the atomic limit, the probability for the transition $\lambda \rightarrow n$ averaged over phase fluctuations is given by
\new{
\begin{eqnarray}
  w_{\lambda \rightarrow n} &=& \langle  a_n^\ast  a_n \rangle_b = \\
\nonumber  &=& \int d\omega \, J(\omega) n_B(\omega) \left| Q_T(E_\lambda + \omega - E_n) \right|^2 \\ &\times& \nonumber
  \frac{|I_{n \lambda}|^2}{\left| E_\lambda + \omega - E_n + i \delta_+ \right|^2}.
\end{eqnarray}
For $\delta_+\to 0$, using 
$|Q_T(E)|^2  | E + i \delta_+|^{-2} \to  2 \pi \tau_+ \delta(E),$}
the transition rate between Andreev states $\lambda\to n$ follows as
\new{
\begin{equation}\label{as:Glmbdn}
  \Gamma_{\lambda \to n} = \frac{w_{\lambda \to n}}{\tau_+} =
  2\pi |I_{n \lambda}|^2 J(\omega) n_B(\omega) \Big|_{\omega = E_n - E_\lambda}.
\end{equation}
}As one may have expected, Eq.~\eqref{as:Glmbdn} differs from the atomic-limit result \eqref{al:Glmbdn} only in the current matrix elements. 
For $\Delta\to \infty$, Eq.~\eqref{as:Ilmbdn} recovers the current matrix element for the atomic limit specified after Eq.~\eqref{al:Ilmbdn}.
Moreover, Eq.~\eqref{as:Glmbdn} also agrees with previous derivations based on the BdG formalism \cite{Zazunov2014,Ackermann2023,Zatsarynna2024}.
One easily checks that the transition rate \eqref{as:Glmbdn} satisfies a detailed balance relation,
\begin{equation}\label{as:detbal}
  \Gamma_{n \to \lambda} = e^{-(E_\lambda-E_n)/T_b} \Gamma_{\lambda \to n} ,
\end{equation}
which connects the rates for forward and backward transitions
as required for equilibrium fluctuations. In addition, particle-hole symmetry  (in particular, $I_{n \lambda} = - I_{\bar n \bar \lambda}^\ast$) implies the symmetry relation
$\Gamma_{n \to \lambda} = \Gamma_{\bar \lambda \to \bar n}$  \cite{Zatsarynna2024}.

Before turning to the transition rates between ABSs and continuum states, 
let us briefly discuss the spectral function 
$S(\omega) =i [ G^R(\omega) - G^A(\omega)]$ of the junction in the absence of phase fluctuations.
Due to the particle-hole symmetry relation \eqref{eom:PHsymm}, the GFs satisfy the relations
\begin{equation}
  G^{R/A}(\omega) = - \tau_x \left[ G^{R/A}(-\omega) \right]^\ast \tau_x,
\end{equation}
and hence the spectral function obeys the constraint
\begin{equation}\label{sf:PHsymm}
  S(-\omega) = \tau_x S^T(\omega) \tau_x.
\end{equation}
Note that $S(\omega)$ is Hermitian for all $\omega$ and thus can be diagonalized,
\begin{equation}\label{sf:Sdiag}
  S(\omega) \to {\rm diag}\left[
  \left( \begin{array}{cc}
    \rho_1(\omega) & 0 \\
    0 & \rho_{\bar 1}(\omega)
  \end{array} \right), \cdots,
  \left( \begin{array}{cc}
    \rho_{2 \ell}(\omega) & 0 \\
    0 & \rho_{\bar{2 \ell}}(\omega)
  \end{array} \right)
  \right],
\end{equation}
where the $2 \times 2$ matrices act in Nambu space. 
The spectral eigenvalues $\rho_n(\omega) \geq 0$ with $n \in \{ \nu, \bar \nu \}$ correspond to the density of states for quasiparticles with energy $\omega$. They 
are related by particle-hole symmetry, see Eq.~\eqref{sf:PHsymm}, according to  $\rho_\nu(\omega) = \rho_{\bar \nu}(-\omega).$
In particular, in the subgap region, $h(\omega)$ is Hermitian and $S(\omega)$ corresponds to a set of $\delta$-function peaks located at the ABS energies,
\begin{equation}\label{sf:SforAL}
  S(\omega) \Big|_{ |\omega| < \Delta} = 
  2 \pi \sum_{\nu = 1}^{2\ell} \left[
  \eta_\nu {\eta}^\dagger_\nu \delta(\omega - E_\nu) + \eta_{\bar \nu} {\eta}^\dagger_{\bar \nu} \delta(\omega + E_\nu)  \right].
\end{equation}

\subsubsection{Transition rates between ABSs and continuum states}

Next, to compute transition rates connecting ABSs to the continuum sector,
we first recall that after integrating out the superconducting leads, the continuum quasiparticles are encoded in $G^{R/A}(\omega)$. However, they
cannot be described by eigenstates of $h(\omega) \neq h^\dagger(\omega)$ for $|\omega| > \Delta$. To calculate the transition rates, we identify the continuum modes via the eigenstates of the spectral function $S(\omega)$. Based on Eq.~\eqref{sf:Sdiag}, 
\begin{equation}\label{sf:xin}
  S(\omega) \xi_n(\omega) = 2 \pi \rho_n(\omega) \xi_n(\omega),\quad
  {\xi}^\dagger_n(\omega) \xi_{n'}(\omega) = \delta_{nn'},
\end{equation}
where the $\xi_n(\omega)$ are $4\ell$-dimensional multispinors in level-Nambu space 
with $n \in \{ \nu,\bar{\nu} \}$ and $\nu \in \{ 1, \ldots, 2\ell\}$. 
Using the completeness of the basis $\{\xi_n(\omega)\}$ for given $\omega$, we write
\begin{equation}
  S(\omega) = 2 \pi \sum_{n \in \{ \nu,\bar{\nu} \}}\rho_n(\omega) \xi_n(\omega) \xi^\dagger_n(\omega),
\end{equation}
which generalizes Eq.~\eqref{sf:SforAL} to arbitrary $\omega$, including continuum states with $|\omega| > \Delta$.

To proceed, we expand the continuum wave function $\tilde{\Psi}(t)$, see Eq.~\eqref{tr:Psi1}, in the $S(\omega)$ eigenstate basis \eqref{sf:xin},
\begin{equation}
  \tilde{\Psi}(t) = \int_{|E| > \Delta} dE \sum_{n \in \{ \nu,\bar{\nu} \}} \tilde a_n(E) \rho_n(E) \xi_n(E) e^{- i E t},
\end{equation}
where $\tilde a_n(E)$ are the corresponding probability amplitudes for continuum modes at long times $t=T$. 
From Eq.~\eqref{tr:Psi1}, we now find
\new{
\begin{eqnarray}\nonumber 
  \tilde a_n(E) &=& \frac{1}{\pi \rho_n(E)} \int_{0}^{T} dt e^{i E t} {\xi}^\dagger_n(E) \Psi^{(1)}_\lambda(t)\\
  &=& \frac{1}{\pi \rho_n(E)} \int \frac{d\omega}{2\pi} \, \frac{\phi_\omega}{2} \,
  {\cal Q}_T^\ast(E_\lambda + \omega - E)  \\ 
  \nonumber &\times& {\xi}^\dagger_n(E) G^R(E_\lambda + \omega)\left[
  \tau_z {\cal I}(E_\lambda) - {\cal I}(E_\lambda + \omega) \tau_z
  \right] \eta_\lambda,
\end{eqnarray}
}with $T \rightarrow \infty$ and \new{${\cal Q}_T(E) = T Q_T(E)$}, see Eq.~\eqref{as:QT}.
Using  the completeness identity $\sum_{k \in \{ \nu,\bar{\nu} \}} \xi_k(E)  \xi^\dagger_k(E) = 1$ 
and the Lehmann representation for $G^R(\omega)$,
\begin{equation}
  G^R(\omega) = \int \frac{d z}{2\pi} \frac{S(z)}{\omega - z + i0^+},
\end{equation}
we obtain \new{
\begin{eqnarray} \nonumber
  \tilde a_n(E) &=& \frac{1}{\pi \rho_n(E)} \int \frac{d\omega}{2\pi} \, \frac{\phi_\omega}{2} \,
  {\cal Q}_T^\ast(E_\lambda + \omega - E) \\   &\times& \sum_{k \in \{ \nu,\bar{\nu} \}}
  G^R_{nk}(E) \tilde I_{k \lambda}(E),
\end{eqnarray}
where
\begin{equation}\label{cs:tldI}
  \tilde I_{k \lambda}(E) = {\xi}^\dagger_k(E) \left[
  \tau_z {\cal I}(E_\lambda) - {\cal I}(E_\lambda + \omega) \tau_z \right] \eta_\lambda,
\end{equation}
and
\begin{eqnarray}\label{cs:GRnk}
  G^R_{nk}(E) &=& {\xi}^\dagger_n(E) G^R(E_\lambda + \omega) \xi_k(E) \\  \nonumber
  &=& \sum_{m \in \{ \nu,\bar{\nu} \}} \int \frac{d z \, \rho_m(z)}{E_\lambda + \omega - z + i0^+} \\ \nonumber &\times&
  \left( {\xi}^\dagger_n(E) \xi_m(z) \right) \left( {\xi}^\dagger_m(z) \xi_k(E) \right).
\end{eqnarray}
}Following the same procedure as before, the probability for the transition $\lambda \to (E, n)$ between the ABS with index $\lambda$ and the continuum mode with index $n$ at energy $E$, averaged over phase fluctuations, is given by
\new{
\begin{eqnarray}
 && w_{\lambda \rightarrow (E, n)} = \langle  \tilde a_n^\ast(E) \tilde a_n(E) \rangle_b \\ &&= \nonumber
   \frac{1}{\pi^2 \rho_n^2(E)}
   \int d\omega \, J(\omega) n_B(\omega) \left| {\cal Q}_T(E_\lambda + \omega - E) \right|^2 \\ 
   \nonumber &&\times \quad
  \Big| \sum_{k \in \{ \nu,\bar{\nu} \}} G^R_{nk}(E) \tilde I_{k \lambda}(E) \Big|^2.
\end{eqnarray}
Taking into account that
\begin{equation}
  \lim_{T \to \infty} \frac{d}{dT} \, \left| {\cal Q}_T(E) \right|^2 =
  \lim_{T \to \infty} \frac{\sin(E T)}{E / 2} = 2 \pi \delta(E),
\end{equation}
 the transition rate follows as 
\begin{eqnarray}\nonumber
  \Gamma_{\lambda \rightarrow (E, n)} &=& \lim_{T \to \infty} \frac{d}{dT} \, w_{\lambda \rightarrow (E, n)} \\
  &=& \nonumber
  \frac{2}{\pi \rho_n^2(E)} \int d\omega \, J(\omega) n_B(\omega) \delta(E_\lambda + \omega - E) \\ 
  &\times&
  \Big| \sum_{k \in \{ \nu,\bar{\nu} \}} G^R_{nk}(E) \tilde I_{k \lambda}(E) \Big|^2,
\end{eqnarray}
with the current matrix elements, see Eqs.~\eqref{cs:tldI} and \eqref{cs:GRnk} with $E_\lambda + \omega = E$,}
\begin{equation}
  \tilde I_{k \lambda}(E) = {\xi}^\dagger_k(E) \left[
  \tau_z {\cal I}(E_\lambda) - {\cal I}(E) \tau_z \right] \eta_\lambda,
\end{equation}
and using Eq.~\eqref{cs:GRnk}.
Assuming that $\rho_m(z)$ is a smooth function of $z$ for all $m$, one can discard the principal value integrals in Eq.~\eqref{cs:GRnk}, which yields
\begin{equation}
  G^R_{nk}(E) = - i \pi \delta_{n k} \rho_n(E),
\end{equation}
and hence, see Eq.~\eqref{as:Glmbdn}, \new{
\begin{equation}\label{cs:GammaEn}
  \Gamma_{\lambda \to (E, n)} = 2 \pi \int d\omega \, J(\omega) n_B(\omega)
  \delta(E_\lambda + \omega - E) \Big| \tilde I_{n \lambda}(E) \Big|^2.
\end{equation}
}

The total quasiparticle escape rate out of the ABS sector is obtained by summing over all partial rates \eqref{cs:GammaEn}, 
\begin{equation}\label{cs:GammaEn_integrated}
  \Gamma_\lambda^{\rm out} = \int_{|E| > \Delta} dE \sum_{n \in \{ \nu,\bar{\nu} \}} \rho_n(E) \left[
  1 - n_F(E) \right] \Gamma_{\lambda \rightarrow (E, n)},
\end{equation}
where $n_F(E)$ is the Fermi-Dirac function with temperature $T_{\rm qp}$.
We arrive at \new{
\begin{eqnarray}\nonumber
  \Gamma_\lambda^{\rm out} &=& 2 \pi \int d\omega \, J(\omega) n_B(\omega)
   \sum_{n \in \{ \nu,\bar{\nu} \}} \Theta(|E| - \Delta) \\ \label{gout}
  &\times & \rho_n(E) \left[
  1 - n_F(E) \right] \Big| \tilde I_{n \lambda}(E) \Big|^2 \, \Bigg|_{E = E_\lambda + \omega}.
  \label{eq:conti_out}
\end{eqnarray}
The reverse rate follows from the detailed balance relation \eqref{as:detbal},
\begin{eqnarray} \nonumber
  \Gamma_{(E, n) \to \lambda} &=&
  2 \pi \int d\omega \, J(\omega) \left[ n_B(\omega) + 1 \right] \\ &\times&
  \delta(E_\lambda + \omega - E) \Big| \tilde I_{n \lambda}(E) \Big|^2,
\end{eqnarray}
such that the total transition rate from the continuum band to the ABS with index $\lambda$ is
\begin{eqnarray}\nonumber
  \Gamma_\lambda^{\rm in} &=& 2 \pi \int d\omega \, J(\omega) \left[ n_B(\omega) + 1 \right]
   \sum_{n \in \{ \nu,\bar{\nu} \}} \Theta(|E| - \Delta) \\ \label{gin}
  &\times& \rho_n(E) n_F(E) \Big| \tilde I_{n \lambda}(E) \Big|^2 \,\Bigg|_{E = E_\lambda + \omega}.
  \label{eq:conti_in}
\end{eqnarray}
}The above expressions are consistent with those derived by the BdG approach in Refs.~\cite{Ackermann2023,Zatsarynna2024}, but allow one to treat the case of a Josephson dot without solving the full BdG scattering problem. Let us emphasize that Eqs.~\eqref{eom:ALScrheq} and \eqref{sf:xin}
 contain just matrices in dot-level space whereas the solution of the full BdG equation requires, in addition, the matching of spatially dependent wave functions of the leads. In our approach, the spatially dependent structure of the wave functions is efficiently encapsulated by the boundary GFs in Eq.~\eqref{gfleads}, thereby circumventing 
 an explicit construction of the BdG solution.

\subsection{Lindblad master equation}\label{sec2d}

We finally summarize the Lindblad master equation \cite{breuer2007theory} describing the dynamics of the ABS sector within the above approximations, see also Ref.~\cite{Zatsarynna2024}. 
In second-quantized notation and carefully taking into account double-counting effects,
the reduced density matrix $\rho_A(t)$ describing the state dynamics in the ABS sector obeys the Lindblad equation
\begin{eqnarray}  \nonumber
\partial_t \rho_A &= &-i\sum_\lambda E_\lambda [a_\lambda^\dagger a_\lambda^{}, \rho_A] +  \sum_{\lambda,\lambda'} 
\Biggl( \Gamma_{\lambda'\to \lambda} \,{\cal L}\left[a_\lambda^\dagger a_{\lambda'}^{}\right]\rho_A   +  \\ &+&
\frac12\left(\Gamma_{\lambda'\to \bar\lambda}\, {\cal L} \left[ a_{\lambda}^{ } a_{\lambda'}^{ }\right] \rho_A
+ \Gamma_{\bar\lambda'\to \lambda} \,{\cal L} \left[ a_{\lambda}^{\dagger} a_{\lambda'}^{\dagger}\right] \rho_A\right)
    \Biggr) \nonumber \\ \label{lindbladA}
 &+& \sum_{\lambda} \left(\Gamma_\lambda^{\rm in} \, \mathcal{L}[a^{\dagger}_{\lambda}] \rho_A + 
  \Gamma_\lambda^{\rm out} \, \mathcal{L}[a^{}_{\lambda}] \rho_A  \right),
\end{eqnarray}
with the dissipator superoperator ${\cal L}[a]\rho_A = a\rho_A a^\dagger - \frac12 \{ a^\dagger a,
\rho_A\}$ and the anticommutator $\{ \cdot,\cdot \}$.  The fermion annihilation operator $a_\lambda$ refers to  the
 ABS with energy $E_\lambda>0$. To avoid double counting, 
 summations over $\lambda$ variables extend only over positive-energy states.
 Negative-energy states are described by the particle-hole symmetry relations $E_{\bar\lambda}=-E_\lambda$ and $a_{\bar\lambda}=a_\lambda^\dagger$.
The transition rates $\Gamma_{n\to n'}$ between ABSs (with energies of either sign) are specified in Eq.~\eqref{as:Glmbdn},
and the rates $\Gamma_\lambda^{\rm out/in}$ connecting ABSs to the continuum sector are given by
Eqs.~\eqref{gout} and \eqref{gin}, respectively.

\section{Quantum Mpemba Effect}\label{sec3}

In this section, we apply the general formalism presented in Sec.~\ref{sec2} to a study of the QME 
in the setup of Fig.~\ref{fig1}.  In Sec.~\ref{sec3a}, we show that already the simplest case of a phase-quenched
short single-channel junction without SOI and/or Zeeman field gives rise to different types of QMEs. 
The short-junction case also allows for analytical progress in the limit $\Gamma\gg \Delta$, where $\Gamma$ is
the hybridization between the dot level and the BCS leads.    
 In Sec.~\ref{sec3b}, we turn to  an intermediate-length junction with SOI and 
Zeeman field, where we present numerical results on the QME.  
For simplicity, we always put the chemical potential $\mu=0$ in the quantum dot region.

As discussed in Sec.~\ref{sec2b}, we assume that the transition rates induced by the electromagnetic environment (bosonic bath)
dominate over phonon-induced rates, and neglect phonon-related effects in what follows.  Many-body states 
in the Andreev sector then equilibrate according to transition rates determined by the bosonic temperature $T_b$ characterizing the electromagnetic environment.  In addition, Andreev states are also
coupled to the fermionic bath corresponding to BCS quasiparticles in the superconducting leads, and thus these transition rates also depend on the temperature $T_{\rm qp}$ of the fermionic bath. We 
assume that quasiparticles are in thermal equilibrium and focus on the 
case $T_{\rm qp}\ge T_b$.

\subsection{Short-junction case}\label{sec3a}

We begin with short Josephson junctions without SOI and Zeeman field.
We assume that $V(x)$ in Eq.~\eqref{md:hath} is a hard-wall potential of short length $L\ll v_{\rm F}/\Delta$.
 Below, $\epsilon=\pi^2/(2m_x L^2)$ is the bare dot level energy and, assuming spin-independent tunnel junctions, the hybridizations to the left and right leads are given by $\Gamma_j = 2t_j^2/L$, see Eq.~\eqref{hybrid}. Assuming symmetric tunnel couplings in Eq.~\eqref{md:Htunx}, $t_1=t_2=t_0$, we obtain $\Gamma_L=\Gamma_R\equiv \Gamma/2 = 2t_0^2/L$.  
The junction then allows only for a single (positive energy) spin-degenerate ABS with energy $0<E_1(\phi_0)<\Delta$. As a consequence, there are four many-body Andreev states, $\left|n\right\rangle \in\left\{ \left|0\right\rangle ,\left|\uparrow\right\rangle ,\left|\downarrow\right\rangle ,\left|\uparrow\downarrow\right\rangle \right\}$, which correspond to an empty, singly occupied (with either spin up or down, $\sigma\in \{\uparrow,\downarrow\}$), or doubly occupied ABS level, respectively. 

\begin{figure}
\begin{center}
    \includegraphics[width=0.46\textwidth]{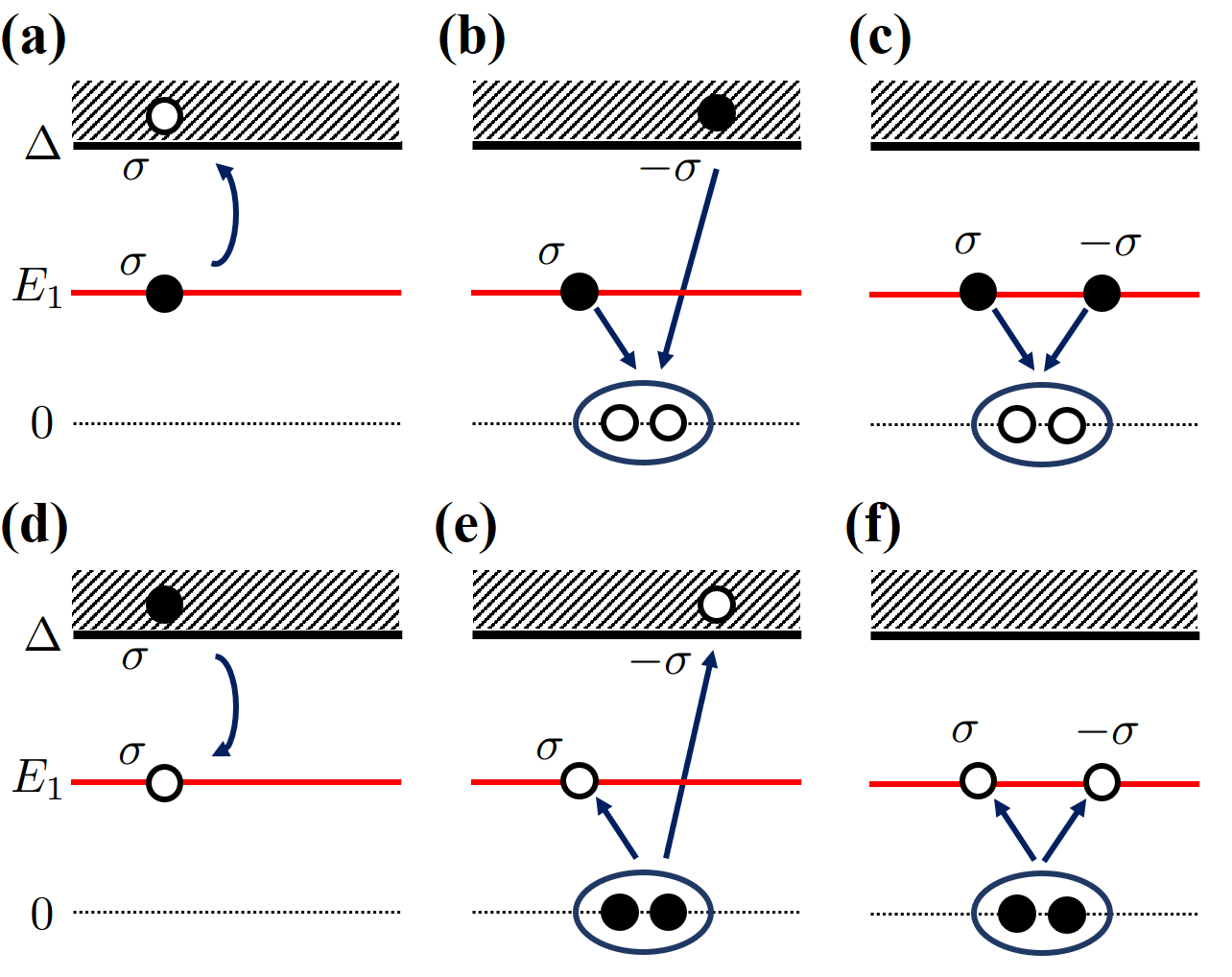}
    \caption{Schematic illustration of the six transition rates contributing to the off-diagonal matrix elements in Eq.~\eqref{pauli}. In all panels, black (open) dots refer to the initial (final) population. Encircled double dots indicate Cooper pairs.  Panel (a) shows the
    transition of a quasiparticle of spin $\sigma\in\{\uparrow,\downarrow\}$ from the ABS energy $E_1(\phi_0)$ to a continuum level with energy $E\ge\Delta$. Panel (d) shows the reverse process. 
    Panels (b) and (c) show processes involving fermion pair annihilation, with and without the contribution of a continuum quasiparticle, respectively. Panels (e) and (f) show the reverse processes, where fermion pairs are created.
    In panels (b,c) and (e,f), the quasiparticle spins must be anti-aligned.  }
    \label{fig2}
\end{center}
\end{figure}

As first step, we then project the Lindblad equation \eqref{lindbladA} for the Andreev quantum state $\rho_{A}(t)$ into the many-body basis $\{|n\rangle\}$. The diagonal elements of $\rho_{A}(t)$ define the respective occupation probabilities, 
$P_{\left|n\right\rangle}(t)=\left\langle n\right|\rho_{A}(t)\left|n\right\rangle$ with  $\sum_{n}P_{\left|n\right\rangle }(t)=1$, which are summarized  in the vector 
\begin{equation}
    \mathbf{P}(t)=\left(P_{\left|0\right\rangle }(t),P_{\left|\uparrow\right\rangle }(t),P_{\left|\downarrow\right\rangle }(t),P_{\left|\uparrow\downarrow\right\rangle }(t)\right)^{T}. 
\label{pauli_master}
\end{equation}
For our model, off-diagonal elements of $\rho_A(t)$ (``coherences'') decouple from $\mathbf{P}(t)$ which in turn obeys a Pauli master equation,
\begin{equation}
\dot{\mathbf{P}}(t)=\mathbf{M}\,\mathbf{P}(t),
\end{equation}
with the matrix 
\begin{equation}
\mathbf{M}=\left(\begin{array}{cccc}
-M_{\left|0\right\rangle } & \Gamma_a^{-}+\Gamma_{b}^{-} & \Gamma_{a}^{-}+\Gamma_{b}^{-} & \Gamma_{c}^{-}\\
\Gamma_{a}^{+}+\Gamma_{b}^{+} & -M_{\left|\uparrow\right\rangle } & 0 & \Gamma_{a}^{-}+\Gamma_{b}^{-}\\
\Gamma_{a}^{+}+\Gamma_{b}^{+} & 0 & -M_{\left|\downarrow\right\rangle } & \Gamma_{a}^{-}+\Gamma_b^{-}\\
\Gamma_{c}^{+} & \Gamma_{a}^{+}+\Gamma_{b}^{+} & \Gamma_{a}^{+}+\Gamma_{b}^{+} & -M_{\left|\uparrow\downarrow\right\rangle }
\end{array}\right).
\label{pauli}
\end{equation}
The off-diagonal matrix elements of $\mathbf{M}$ contain transition rates
for the physical processes  illustrated in Fig.~\ref{fig2}. These rates follow from Sec.~\ref{sec2c} and are explained in detail below.
Probability conservation implies that the quantities $M_{\left|n\right\rangle }$ in Eq.~\eqref{pauli} are equal to the sum of the off-diagonal elements in the corresponding columns.

\subsubsection{Case $\Gamma\gg \Delta$}\label{sec3a1}

We first describe an analytical approach for identifying the QME for a single-level dot with large hybridization to the superconducting leads, $\Gamma\gg \Delta$.  We go beyond this restriction in  Sec.~\ref{sec3a3} by performing  numerical calculations.
For $\Gamma\gg \Delta$, the quantum dot model in Sec.~\ref{sec2a} implies the well-known ABS dispersion relation \cite{Beenakker1991,Furusaki1991,Nazarov2009}
\begin{equation}\label{ABS1}
    E_1(\phi_0) \simeq \Delta \sqrt{1-{\cal T}\sin^2(\phi_0/2)},
\end{equation}
where the normal-state transmission probability of the transport channel is given by 
\begin{equation} \label{transparency_short}
    {\cal T} = \frac{1}{1+(\epsilon/\Gamma)^2}.
\end{equation}

The rate $\Gamma_{a}^{-}$ shown in Fig.~\ref{fig2}(a) describes the transition of a quasiparticle with spin $\sigma$
(the rate is independent of $\sigma$) from the ABS to the quasiparticle continuum. Here and in what follows, we label rates associated with processes that decrease (increase) the number of ABS quasiparticles by $\Gamma^-$ ($\Gamma^+$). 
The rate $\Gamma_a^-$ decreases this number by one unit and requires the
absorption of a photon with energy $\delta E_a=E-E_1(\phi_0)$, where $E \ge \Delta$ is the energy of the continuum quasiparticle, together with the possibility to allocate the quasiparticle in the continuum. 
Integrating over the allowed continuum energies, the transition rate follows from Eq.~\eqref{cs:GammaEn_integrated} by limiting the integral to $E>\Delta$, see also Ref.~\cite{Zatsarynna2024},
\begin{equation}\label{g1t}
\Gamma_{a}^{-}=\int_{E > \Delta} dE\, \gamma_{a}(E,\phi_0)\, J\left(\delta E_a \right) n_B\left(\delta E_a \right) \left[1-n_F\left(E\right)\right].
\end{equation}
We recall that $n_B$ is the Bose-Einstein function for the temperature $T_b$, while $n_F$ is the Fermi-Dirac function for the temperature $T_{\rm qp}$.
Applying the formalism detailed in Sec.~\ref{sec2c} in the limit $\Gamma\gg \Delta$, we find
\begin{equation}
\gamma_{a}(E,\phi_0)=\frac{\Gamma\nu(E)}{2}\left( 1+ \frac{\Delta^2 \cos^2{(\phi_0/2)}}{E E_1(\phi_0)}\right),
\end{equation}
where the superconducting density of states is encoded by the function 
\begin{equation}\label{sdos}
    \nu(E)=\Theta(|E|-\Delta)\frac{|E|}{\sqrt{E^2-\Delta^2}}.
\end{equation}
The complementary process to Eq.~\eqref{g1t} is a transition from the continuum
to the ABS, see Fig.~\ref{fig2}(d). The corresponding rate is proportional to the probability to
encounter a continuum quasiparticle times the amplitude for spontaneous or stimulated photon
emission, 
\begin{equation}
\Gamma_{a}^{+}=\int_{E > \Delta} dE\, \gamma_{a}(E,\phi_0)\, J\left(\delta E_a\right) \left[1+n_B\left(\delta E_a\right)\right] n_F\left(E\right),
\end{equation} 
again with $\delta E_a=E-E_1(\phi_0)$.

Next, the rates $\Gamma_{b}^{\mp}$ encode the creation or annihilation of a Cooper pair by means of an ABS quasiparticle with spin $\sigma$ and a continuum quasiparticle with spin $-\sigma$, see
Figs.~\ref{fig2}(b) and Fig.~\ref{fig2}(e), respectively. (Again, the result is independent of  $\sigma.$)
Cooper pair annihilation comes with the energy cost $\delta E_b=E+E_1(\phi_0)$, and the respective transition rates are encoded in Eq.~\eqref{cs:GammaEn_integrated}, but limiting the integral to $E<-\Delta$ \cite{Zatsarynna2024}. Using the formalism in Sec.~\ref{sec2c} for $\Gamma\gg\Delta$, we can express the rates as
\begin{eqnarray}\nonumber
\Gamma_{b}^{-}&=& \int_{E > \Delta} dE \,\gamma_b(E,\phi_0) \,J(\delta E_b)\left[1+n_B(\delta E_b)\right]n_F(E),
\\ \nonumber
\Gamma_{b}^{+}&=&\int_{E > \Delta} dE \,\gamma_b(E,\phi_0) \,J(\delta E_b)n_B(\delta E_b)\left[1-n_F(E)\right],\\
\end{eqnarray}
with
\begin{equation}
\gamma_{b}(E,\phi_0)=\frac{\Gamma\nu(E)}{2}\left( 1- \frac{\Delta^2 \cos^2{(\phi_0/2)}}{E E_1(\phi_0)}\right).
\end{equation}
Finally, in Fig.~\ref{fig2}(c,f), we illustrate Cooper pair creation and
annihilation processes involving two ABS quasiparticles with opposite spin. Such
processes have an energy cost $\delta E_c=2E_1(\phi_0)$ and change the population of the ABS sector by two units. The corresponding transition rates can be obtained from Eq.~\eqref{as:Glmbdn} with $E_n=-E_\lambda=E_1$, see also Ref.~\cite{Zatsarynna2024}.
They are  given by
\begin{eqnarray}\nonumber
\Gamma_{c}^{-}&=&\gamma_{c}(\phi_0) \, J\left(\delta E_c\right)\, \left[1+n_B\left(\delta E_c\right)\right],\\
\Gamma_{c}^{+}&=&\gamma_{c}(\phi_0) \, J\left(\delta E_c\right)\, n_B\left(\delta E_c\right),
\end{eqnarray}
with
\begin{equation}
\gamma_{c}(\phi_0)=2\pi \left( \frac{\epsilon \Delta \sin{(\phi_0/2)}}{E_1(\phi_0)}\right)^2.
\end{equation}
For each pair of processes above, standard detailed balance conditions \cite{Nazarov2009,Weiss_2012} hold since both the bosonic environment and the fermionic continuum quasiparticles are
separately assumed to be in thermal equilibrium.  
We note that by multiplying $\gamma_{a,b,c}$ by an overall factor, only the total relaxation time is affected, without changing the stationary state $\mathbf{P}_{\rm stat}$
reached at asymptotically long times. The latter state obeys $\dot{\mathbf{P}}_{\rm stat}=0$ and only depends on ratios of transition rates. 

A simplification is possible by exploiting spin degeneracy: the population
difference $P_{\left|\uparrow\right\rangle }(t)-P_{\left|\downarrow\right\rangle }(t)$
decouples from $P_{\left|0\right\rangle }(t)$ and $P_{\left|\uparrow\downarrow\right\rangle }(t)$
and vanishes for $t\to\infty$. 
For the reduced time-dependent population vector 
\begin{equation}\label{prt}
    \mathbf{P}_{r}(t)=\left(P_{\left|0\right\rangle }(t),P_{\left|1\right\rangle }(t),P_{\left|\uparrow\downarrow\right\rangle }(t)\right)^{T},
\end{equation}
where $P_{\left|1\right\rangle }(t)=(P_{\left|\uparrow\right\rangle }(t)+P_{\left|\downarrow\right\rangle }(t))/2$, we thus obtain a reduced Pauli master equation,
\begin{equation}
\dot{\mathbf{P}}_{r}(t)=\mathbf{M}_{r}\mathbf{P}_{r}(t).
\label{pauli_r}
\end{equation}
The normalization condition is here given by
\begin{equation}\label{normr}
 P_{\left|0\right\rangle}(t) +2P_{\left|1\right\rangle}(t) +P_{\left|\uparrow\downarrow\right\rangle}(t) =1.
 \end{equation}
With the above approximations, the matrix $\mathbf{M}_r$ in Eq.~\eqref{pauli_r} follows as
\begin{equation}\label{matrixmr}
\mathbf{M}_{r}=\left(\begin{array}{ccc}
-2\Gamma_{ab}^{+}-\Gamma_{c}^{+} & 2\Gamma_{ab}^{-} & \Gamma_{c}^{-}\\
\Gamma_{ab}^{+} & -\Gamma_{ab}^{-}-\Gamma_{ab}^{+} & \Gamma_{ab}^{-}\\
\Gamma_c^{+} & 2\Gamma_{ab}^{+} & -2\Gamma_{ab}^{-}-\Gamma_{c}^{-}
\end{array}\right),
\end{equation}
with $\Gamma_{ab}^{\pm}=\Gamma_{a}^{\pm}+\Gamma_{b}^{\pm}$.
The stationary solution  of Eq.~\eqref{pauli_r} reached at asymptotically long times is given by 
\begin{equation}\label{pstat}
\mathbf{P}_{r,{\rm stat}} =  \mathcal{N} \left(\begin{array}{c} 2\Gamma_{ab}^{-\, 2}+\Gamma_{ab}^{-}\Gamma_{c}^{-}+\Gamma_{c}^{-}\Gamma_{ab}^{+} \\ 2\Gamma_{ab}^{-}\Gamma_{ab}^{+}+\Gamma_{ab}^{+}\Gamma_{c}^{-}+\Gamma_{c}^{+}\Gamma_{ab}^{-} \\ 2\Gamma_{ab}^{+\, 2}+\Gamma_{ab}^{-}\Gamma_{c}^{+}+\Gamma_{c}^{+}\Gamma_{ab}^{+} \end{array}\right),
\end{equation}
where $\mathcal{N}$ follows by normalization, see Eq.~\eqref{normr}.
As expected for a dissipative master equation satisfying detailed
balance, the stationary solution does not depend on initial conditions
but only on ratios between  transition rates. 
In the following, we focus on how changes of the phase difference $\phi_0$ affect the stationary  populations in Eq.~\eqref{pstat}. We assume  that  phase differences are taken from the interval $\phi_0\in [0,\pi)$ such that 
the phase quench is uniquely related to a quench of the ABS energy $E_1(\phi_0)$, see Eq.~\eqref{ABS1}.
As shown below, by analyzing the $\phi_0$-dependence of $\mathbf{P}_{r,{\rm stat}}$, one can identify parameter regions where a QME is possible when $\phi_0$ is quenched.  We also discuss the corresponding steady-state
current-phase relation (CPR), which for the present case follows in the simple form \cite{Nazarov2009}
\begin{equation}\label{CPR}
I(\phi_0) = 2 \frac{d E_1}{d \phi_0} \left(
P_{|\uparrow \downarrow \rangle,{\rm stat}}(\phi_0) -P_{|0 \rangle,{\rm stat}}(\phi_0) \right).
\end{equation}

\begin{figure}
\begin{center}
    \includegraphics[width=0.43\textwidth]{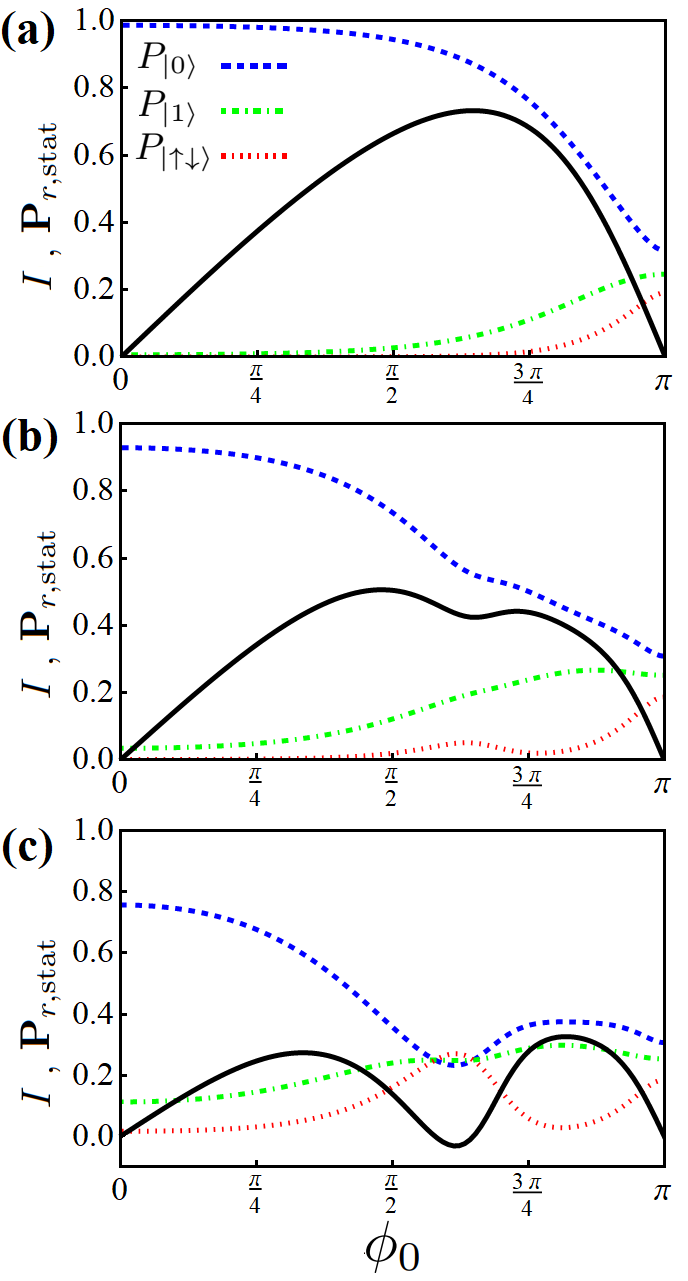}
    \caption{ Steady-state populations $\mathbf{P}_{r,{\rm stat}}$ and CPR vs $\phi_0$, see Eqs.~\eqref{prt}, \eqref{pstat} and \eqref{CPR}, for the 
    many-body Andreev states in a short Josephson junction  coupled to a microwave resonator with spectral density \eqref{jw}.  The current $I$ in the CPR is given in units of $2e\Delta/\hbar$.
    The elements of $\mathbf{P}_{r,{\rm stat}}$ are shown vs $\phi_0$, where $P_{\left|0\right\rangle}$ is indicated by dashed blue curves, $P_{\left|1\right\rangle}$ by dot-dashed green curves, and $P_{\left|\uparrow\downarrow\right\rangle}$  by dotted red curves. The CPR is shown as solid black curve.
    Using units with $\Delta=1$, we 
    use $T_b=0.2$, $\epsilon=0.5$, $\Gamma=10$,  
    $\Omega_e=0.01$, $\eta=0.1$, and $\kappa=0.1$.  
    The corresponding junction transparency is ${\cal T} =0.99$ from Eq.~\eqref{transparency_short}.
    The panels are for (a) $T_{\rm qp}=0.2$, (b) $T_{\rm qp}=0.3$, and (c) $T_{\rm qp}=0.5$.}
    \label{fig3}
\end{center}
\end{figure}

In order to check the robustness of the QME in our setup, we consider two particular choices for the environmental 
spectral density $J(\omega)$. For a microwave-circuit environment with resonance frequency $\Omega_e$,  coupling strength $\kappa$,
and damping constant $\eta$, we use the Lorentzian spectral density \cite{Weiss_2012,Ackermann2023}
\begin{equation}\label{jw}
J(\omega)=\frac{\kappa^{2}\eta}{\pi}\left(\frac{1}{\left(\omega-\Omega_e\right)^{2}+\frac{\eta^{2}}{2}}-\frac{1}{\left(\omega+\Omega_e\right)^{2}+\frac{\eta^{2}}{2}}\right),
\end{equation}
while for an Ohmic environment with damping coefficient $\alpha_d$ and high-frequency cutoff $\omega_c$, we employ \cite{Weiss_2012,Barr_2024}
\begin{equation}\label{jw_ohm}
J(\omega)=\alpha_d \omega e^{-|\omega|/\omega_c}.
\end{equation}
For the examples shown below, we have chosen specific values for the key parameters in these spectral densities. 
However, we have checked that changing, e.g., the values of $\Omega_e$ and/or $\eta$ in Eq.~\eqref{jw} by up to one order of magnitude does not significantly affect the QME. 
Moreover, as discussed above, changing $\kappa$ or $\alpha_d$ implies only a uniform rescaling of all
transition rates, which does not affect the QME.

Qualitatively different behaviors are observed, as shown
in Fig.~\ref{fig3} for a high-transparency junction with the Lorentzian spectral density \eqref{jw}. 
For three different quasiparticle temperatures $T_{\rm qp}\ge T_b$, Fig.~\ref{fig3} shows the components of the steady-state population vector $\mathbf{P}_{r,{\rm stat}}$ as function of the average phase difference $\phi_0$, which in turn is tunable by a magnetic flux \cite{Alvaro2011}. 
For  $T_{\rm qp}=T_b$, see Fig.~\ref{fig3}(a), all components of $\mathbf{P}_{r,{\rm stat}}$
are monotonic functions of $\phi_0$.  Upon raising $T_{\rm qp}$ to an intermediate value, see Fig.~\ref{fig3}(b),  
some components exhibit non-monotonic behavior. In particular, $P_{\left|1\right\rangle}$ peaks around $\phi_0\simeq 0.85\pi$, while $P_{\left|\uparrow\downarrow\right\rangle}$ has a maximum at $\phi_0\simeq 0.65\pi$. 
Finally, for the highest studied value of $T_{\rm qp}$, see Fig.~\ref{fig3}(c), all population components have nearly simultaneous extrema around $\phi_0\simeq 0.6\pi$ and around $\phi_0\simeq 0.8\pi$. 
In Fig.~\ref{fig3}, we also show the steady-state CPR $I(\phi_0)$ for $0\le \phi_0\le \pi$ in the respective panels, 
see Eq.~\eqref{CPR}. 
We observe that upon increasing the ratio $T_{\rm qp}/T_b$, the CPR can feature pronounced minima near those phase
values where one has extremal points in the population components $P_{|\uparrow \downarrow \rangle,{\rm stat}}(\phi_0)$ 
and/or $P_{|0 \rangle,{\rm stat}}(\phi_0)$.  Experimentally, this observation could  help
to identify the interesting regime $T_{\rm qp}>T_b$.  As discussed below, in this regime,
the QME is expected to be realizable.

\begin{figure}
\begin{center}
    \includegraphics[width=0.43\textwidth]{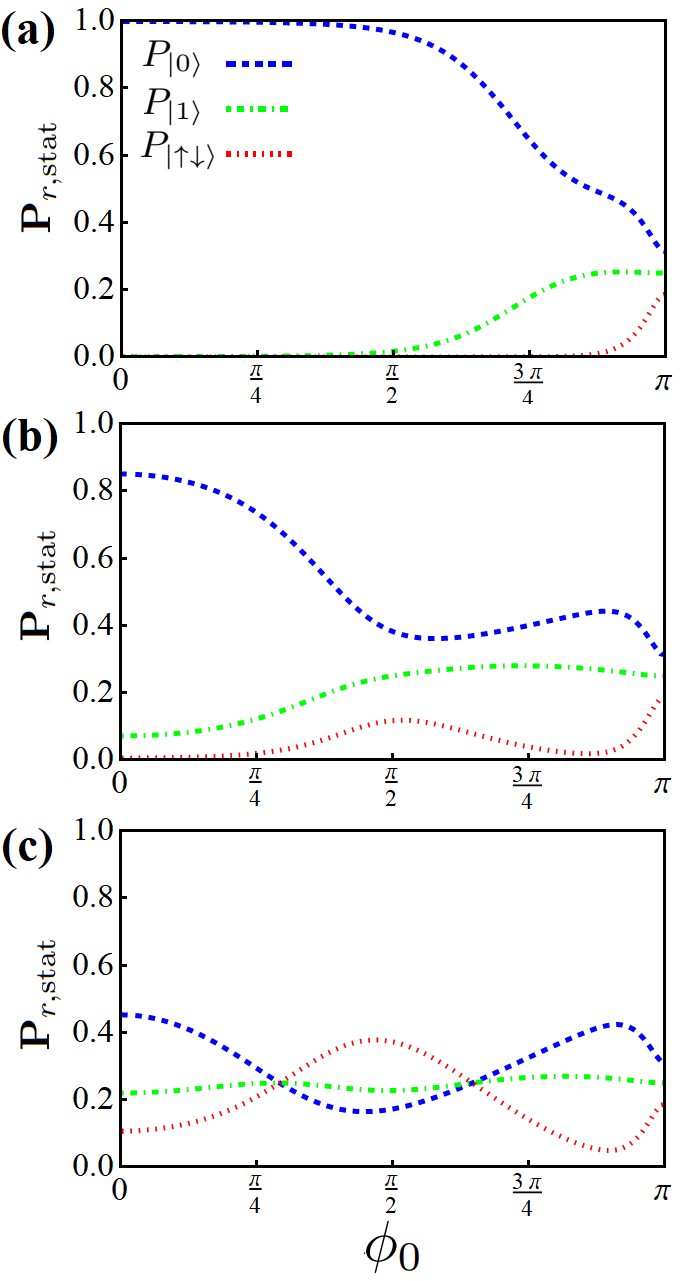}
    \caption{ Steady-state populations $\mathbf{P}_{r,{\rm stat}}$ vs $\phi_0$ for the 
    many-body Andreev states in a short Josephson junction as in Fig.~\ref{fig3} but for the case of an Ohmic environment, see Eq.~\eqref{jw_ohm}.
    Again, $P_{\left|0\right\rangle}$ is indicated by dashed blue curves, $P_{\left|1\right\rangle}$ by dot-dashed green curves, and $P_{\left|\uparrow\downarrow\right\rangle}$  by dotted red curves. 
    Putting $\Delta=1$, we use $T_{b}=0.1$, $\epsilon=0.25$, $\Gamma=10$,   
    $\omega_c=1$, and $\alpha_d=0.1$, such that  
     ${\cal T} =0.99$, see Eq.~\eqref{transparency_short}.
    The panels are for (a) $T_{\rm qp}=0.14$, (b) $T_{\rm qp}=0.3$, and (c) $T_{\rm qp}=0.5$.}
    \label{fig4}
\end{center}
\end{figure}

In Fig.~\ref{fig4}, we show that  qualitatively the same behavior as in Fig.~\ref{fig3} is also encountered for an Ohmic environment described by Eq.~\eqref{jw_ohm}. In particular, 
studying again a high-transparency junction, for a  rather low temperature $T_{\rm qp}$, see Fig.~\ref{fig4}(a), all components of $\mathbf{P}_{r,{\rm stat}}$ remain monotonic functions of $\phi_0$. For an intermediate value of $T_{\rm qp}$, see Fig.~\ref{fig4}(b), the components of $\mathbf{P}_{r,{\rm stat}}$ exhibit 
non-monotonic behavior in some $\phi_0$ regions. In particular, $P_{\left|0\right\rangle}$ and $P_{\left|\uparrow\downarrow\right\rangle}$ share extremal points at $\phi_0\simeq 0.5\pi$ and $\phi_0\simeq 0.9\pi$,
while $P_{\left|1\right\rangle}$ has a maximum for $\phi_0\simeq 0.75\pi$. Finally, for the highest studied value of $T_{\rm qp}$, see Fig.~\ref{fig4}(c), all populations share extremal points around
$\phi_0 \simeq 0.45\pi$ and $\phi_0\simeq 0.85\pi$. 

It is worth stressing that the stationary populations illustrated in Figs.~\ref{fig3} and \ref{fig4} 
depend on which of the six transition rates $\Gamma_{\lambda}^{\pm}$ with $\lambda\in (a,b,c)$  
in Eq.~\eqref{pauli} are dominant. 
For instance, if at least one of the three rates $\Gamma_{\lambda}^{-}$ 
is much larger than all rates $\Gamma_{\lambda'}^{+}$, ABS quasiparticles
tend to be depleted either by hopping into the continuum sector (for $\lambda=a$) or by pair creation processes (for $\lambda=b,c$). One then finds $P_{\left|0\right\rangle} \gg P_{\left|1\right\rangle},P_{\left|\uparrow\downarrow\right\rangle}$, see, e.g., Fig.~\ref{fig3}(a) and Fig.~\ref{fig4}(a) with $\phi_0\approx 0$. On the other hand, if at least one of the rates $\Gamma_{\lambda}^{+}$ exceeds the rates $\Gamma_{\lambda'}^{-}$, excess quasiparticles will be injected into the ABS sector by hopping from the continuum or by processes involving Cooper pair splitting. In this case, $P_{\left|0\right\rangle}$ becomes small. The relative magnitude of $P_{\left|1\right\rangle}$ vs $P_{\left|\uparrow\downarrow\right\rangle}$ is then decided by the parity-preserving rate $\Gamma_c^{+}$ which 
describes the injection of two ABS quasiparticles, and by the rates $\Gamma_{a,b}^{+}$ which increase the ABS population by one unit. In such cases, as shown in Fig.~\ref{fig3}(b,c) and Fig.~\ref{fig4}(b,c) for specific values of $\phi_0$, the stationary populations may exhibit an extremum.
Experimentally, the ratio between transition rates corresponding to different processes can be tuned, for example, via the temperatures $T_b$ and $T_{\rm qp}$, or by changing $\phi_0$.

For the QME protocol, we assume that $\phi_0$ is subject to a rapid quench at time $t=0$.
As we show below, the case in Fig.~\ref{fig3}(a) corresponds to the absence of a QME. However, for
the parameters in Fig.~\ref{fig3}(b,c), two different types of QME as defined in Ref.~\cite{Nava2024} can take place for suitable initial conditions. The same conclusions apply for the corresponding panels in Fig.~\ref{fig4}. Since the overall behavior in Figs.~\ref{fig3} and \ref{fig4} is similar, we expect that the QME is  
robust against changes in the electromagnetic environment. In what follows, we then focus on the Lorentzian spectral density \eqref{jw}.

\subsubsection{QME protocol}

Following Ref.~\cite{Nava2024}, see also Sec.~\ref{sec1},
the protocol for detecting the QME consists of comparing two copies of the system prepared at time $t<0$ 
in the pre-quench stationary states $\mathbf{P}_{r,{\rm stat}}^{(c)}$ and $\mathbf{P}_{r,{\rm stat}}^{(f)}$
 corresponding to the phase differences $\phi_0^{(c)}$ and $\phi_0^{(f)}$, respectively, with all other model parameters kept identical. 
At time $t=0$, for each of these two system copies,
the phase is suddenly quenched to the same post-quench value $\phi_0^{(\rm eq)}$,
where we demand $|\phi_0^{(\rm eq)}-\phi_0^{(c)}|<|\phi_0^{(\rm eq)}-\phi_0^{(f)}|$.
The superscripts $(f)$  vs $(c)$ thus refer to pre-quench values which are ``far'' vs ``close'' to the  
post-quench value, respectively.
Since the ABS dispersion \eqref{ABS1} is symmetric around $\phi_0=\pi$, i.e., $E_1(2\pi-\phi_0) = E_1(\phi_0)$, we take all phases from the interval $\left[0, \pi \right)$. Moreover, since $E_1(\phi_0)$ is a monotonic 
function in this interval, pre- and post-quench ABS energies obey the same ordering.

The corresponding relaxation times $\tau^{(f,c)}$ for reaching the stationary
state $\mathbf{P}_{r,{\rm stat}}^{({\rm eq})}$ at $t\to \infty$ 
are then determined by solving the Pauli master equation \eqref{pauli_r} for $E_1=E_1(\phi_0^{({\rm eq})})$ under 
the initial conditions $\mathbf{P}_r(t=0)=\mathbf{P}_{r,{\rm stat}}^{(f,c)}$.  The QME occurs if $\tau^{(f)}<\tau^{(c)}$.
For the cases shown in Fig.~\ref{fig3}(a) and Fig.~\ref{fig4}(a), the QME is ruled out by the monotonicity of the populations. 
Indeed, with increasing quench amplitude, $\delta \phi_0^{(i\in\{c,f\})}=|\phi_0^{(\rm eq)}-\phi_0^{(i)}|$, 
the distance between the initial $(t=0)$ and final $(t\to \infty$) populations also increases. 
As a consequence, we always find $\tau^{(f)}>\tau^{(c)}$, and thus no QME can occur. 
On the other hand, for the cases shown in Fig.~\ref{fig3}(c) and Fig.~\ref{fig4}(c), increases in $\delta \phi_0^{(i)}$ may cause
a reduction of the effective distance between initial and final populations, resulting in a shorter
relaxation time $\tau$.
Indeed, for certain values of $\delta \phi_0^{(i)}$, we find $\mathbf{P}_{r,{\rm stat}} (\phi_0)\approx \mathbf{P}_{r,{\rm stat}}(\phi_0+\delta \phi_0^{(i)})$. Finally, Fig.~\ref{fig3}(b) and Fig.~\ref{fig4}(b) represent a special intermediate situation discussed below. 

In order to monitor the time evolution of the system and quantitatively detect the QME and its type, 
a proper distance function in Hilbert space must be introduced.  As discussed in Ref.~\cite{Nava2024}, the trace
distance between the time-dependent density matrix $\rho_A(t)$ for $t>0$ and the final steady-state density matrix $\rho^{(\rm eq)}_{A, \rm stat}$ is an admissible choice.
Since in our case $\rho_A(t)$ is effectively diagonal, the trace distance reduces to a norm-1 (\emph{aka} Manhattan) distance \cite{horn2012} between the respective population vectors,
\begin{equation}
\mathcal{D}_{M}\left(\mathbf{P}(t)\right)= \frac{1}{2}\sum_{n}\left|P_{\left|n\right\rangle }(t)-P^{({\rm eq})}_{\left|n\right\rangle ,{\rm stat}}\right|.
\label{norm1}
\end{equation}
This distance function can be obtained experimentally
by measuring ABS populations, which in turn can be achieved, e.g., by microwave spectroscopy, see Refs.~\cite{Janvier2015,Wesdorp2023,Wesdorp2024} and references therein.
We note that $\mathcal{D}_{M}\left(\mathbf{P}(t)\right)\neq\mathcal{D}_{M}\left(\mathbf{P}_r(t)\right)$, since the second component of $\mathbf{P}_r$ needs to be counted twice in Eq.~\eqref{norm1}, see Eq.~\eqref{prt}. 
We can then define the relaxation time $\tau$ by the condition $\mathcal{D}_{M}\left(\mathbf{P}(t)\right)<\epsilon_c$, where $\epsilon_c\ll1$ is a small but  finite accuracy cutoff value. 
The cutoff value is reached for $t=\tau$.
This procedure is necessary since for the relaxation dynamics described by Eq.~\eqref{lindbladA}, there is no phase transition and the true stationary solution is reached only in the asymptotic limit $t\rightarrow\infty$ \cite{Lu2017, Nava2024}. In the following, we set $\epsilon_c=10^{-4}$. However, we have checked that the presence or absence of the QME is robust under variations of the value of $\epsilon_c$.

We next note that Eq.~\eqref{pauli_r} admits the solution
\begin{equation}
\mathbf{P}_{r}(t)=\mathbf{P}^{({\rm eq})}_{r, {\rm stat}}+\sum_{k=1,2} c_k {\bf P}_k e^{\lambda_k t},
\label{pauli_sol}
\end{equation}
where $\lambda_k$ and ${\bf P}_k$ denote the two real-valued negative eigenvalues (we assume $\lambda_2<\lambda_1<0$) and the corresponding right eigenvectors of the matrix ${\bf M}_{r}$ in Eq.~\eqref{matrixmr}, respectively. The coefficients $c_k$ are determined by the initial condition at $t=0$, and 
$\mathbf{P}^{({\rm eq})}_{r, {\rm stat}}$ is given by Eq.~\eqref{pstat} with $\phi_0=\phi_0^{({\rm eq})}$.
In the asymptotic long-time limit, Eq.~\eqref{norm1} reduces to
\begin{equation}\label{a1def}
\mathcal{D}_{M}\left(\mathbf{P}(t)\right)=\frac{1}{2} \mathcal{A}_1 e^{\lambda_1 t},\quad 
\mathcal{A}_1=\left|c_1\right|\sum_n \left|P_{1,\left|n\right\rangle}\right|.
\end{equation}
Here, the $P_{1,|n\rangle}$ are the components of the vector ${\bf P}_k$ in Eq.~\eqref{pauli_sol} with $|\uparrow\rangle=|\downarrow\rangle\equiv |1\rangle$, where we consider only initial conditions with $P_{\left|\uparrow\right\rangle }(0)=P_{\left|\downarrow\right\rangle }(0)$.  The system therefore relaxes to the final stationary state according to an exponential law with the rate $|\lambda_1|$. 
In our quench protocol, this rate only depends on $\phi_0^{({\rm eq})}$. 
The onset of the QME is thus fully determined by the value of $\mathcal{A}_1$, which in turn depends on the 
initial condition $\mathbf{P}_{r}(0)$. Indeed, for given accuracy cutoff $\epsilon_c$, the relaxation times $\tau^{(c,f)}$ 
follow in the form
\begin{equation}
\tau^{(c,f)}\simeq | \lambda_1 |^{-1} \ln\left( \frac{\mathcal{A}_1^{(c,f)}}{2\epsilon_c}\right).
\label{rel_time}
\end{equation}
For $\mathcal{A}_1=0$, see Eq.~\eqref{a1def}, the relaxation dynamics becomes exponentially accelerated with the larger rate $|\lambda_2|$ \cite{Carollo_2021}. The corresponding
relaxation time is computed as in Eq.~\eqref{rel_time} but with $\mathcal{A}_1\to \mathcal{A}_2$ and $\lambda_1\to \lambda_2$.
 
The QME is realized if $\mathcal{A}_1^{(f)}<\mathcal{A}_1^{(c)}$ such that  
\begin{equation}
\mathcal{D}_{M}\left(\mathbf{P}^{(f)}(t)\right)<\mathcal{D}_{M}\left(\mathbf{P}^{(c)}(t)\right)
\label{condition_mpemba}
\end{equation}
is satisfied at long times. 
One can then distinguish type-I and type-II QMEs \cite{Nava2024}.
For the type-I QME, Eq.~\eqref{condition_mpemba} must hold for all $t>0$.  On the other hand, for the type-II QME, 
Eq.~\eqref{condition_mpemba} holds only for $t>t^*$, where $t^*$ denotes a critical time where 
$\mathcal{D}_{M}\left(\mathbf{P}^{(f)}(t)\right)$ and $\mathcal{D}_{M}\left(\mathbf{P}^{(c)}(t)\right)$ intersect. Hence  the condition
\begin{equation}
\mathcal{D}_{M}\left(\mathbf{P}^{(f)}(0)\right)< \mathcal{D}_{M}\left(\mathbf{P}^{(c)}(0)\right)
\label{condition_mpembaI}
\end{equation}
yields a type-I QME, while  
\begin{equation}
\mathcal{D}_{M}\left(\mathbf{P}^{(f)}(0)\right)> \mathcal{D}_{M}\left(\mathbf{P}^{(c)}(0)\right)
\label{condition_mpembaII}
\end{equation}
together with the existence of a critical time $t^\ast$ 
identifies a type-II QME. 

The monotonic $\phi_0$-dependence of the stationary populations for the parameters 
in Fig.~\ref{fig3}(a) or Fig.~\ref{fig4}(a) implies that Eq.~\eqref{condition_mpemba} cannot be satisfied for any choice of $( \phi_0^{(c)}, \phi_0^{(f)},\phi_0^{({\rm eq})})$. Hence the QME is ruled out in these cases. 
However, if all components of $\mathbf{P}_{r}(t)$ have extrema at nearly the same values of $\phi_0$, 
as observed in Fig.~\ref{fig3}(c) and Fig.~\ref{fig4}(c), it is possible to choose $\phi_0^{(c)}$ and $\phi_0^{(f)}$ such that for all $n$,
\begin{equation}
    \left|P_{\left|n\right\rangle }^{(f)}(t)-P^{({\rm eq})}_{\left|n\right\rangle ,{\rm stat}}\right|<\left|P_{\left|n\right\rangle }^{(c)}(t)-P^{({\rm eq})}_{\left|n\right\rangle ,{\rm stat}}\right|.
\end{equation}
As a consequence, Eqs.~\eqref{condition_mpemba} and \eqref{condition_mpembaI} are both satisfied, and a
type-I QME takes place. 
The onset of the more elusive type-II QME requires an intermediate scenario 
such as the one shown in Fig.~\ref{fig3}(b) or Fig.~\ref{fig4}(b). In this case, for some interval of $\phi_0$, 
a subset of the populations have a monotonic dependence while others exhibit an extremal point. 
Due to this feature, Eq.~\eqref{condition_mpembaII} can hold.  In particular, if the monotonic populations reach equilibrium faster than the non-monotonic ones, Eq.~\eqref{condition_mpemba} can be restored at some time $t=t^*$.
To conclude,  both types of QMEs allowed in open quantum systems \cite{Nava2024} could be realized in a highly transparent phase-quenched short Josephson junction.

\begin{figure}
\begin{center}
    \includegraphics[width=0.5\textwidth]{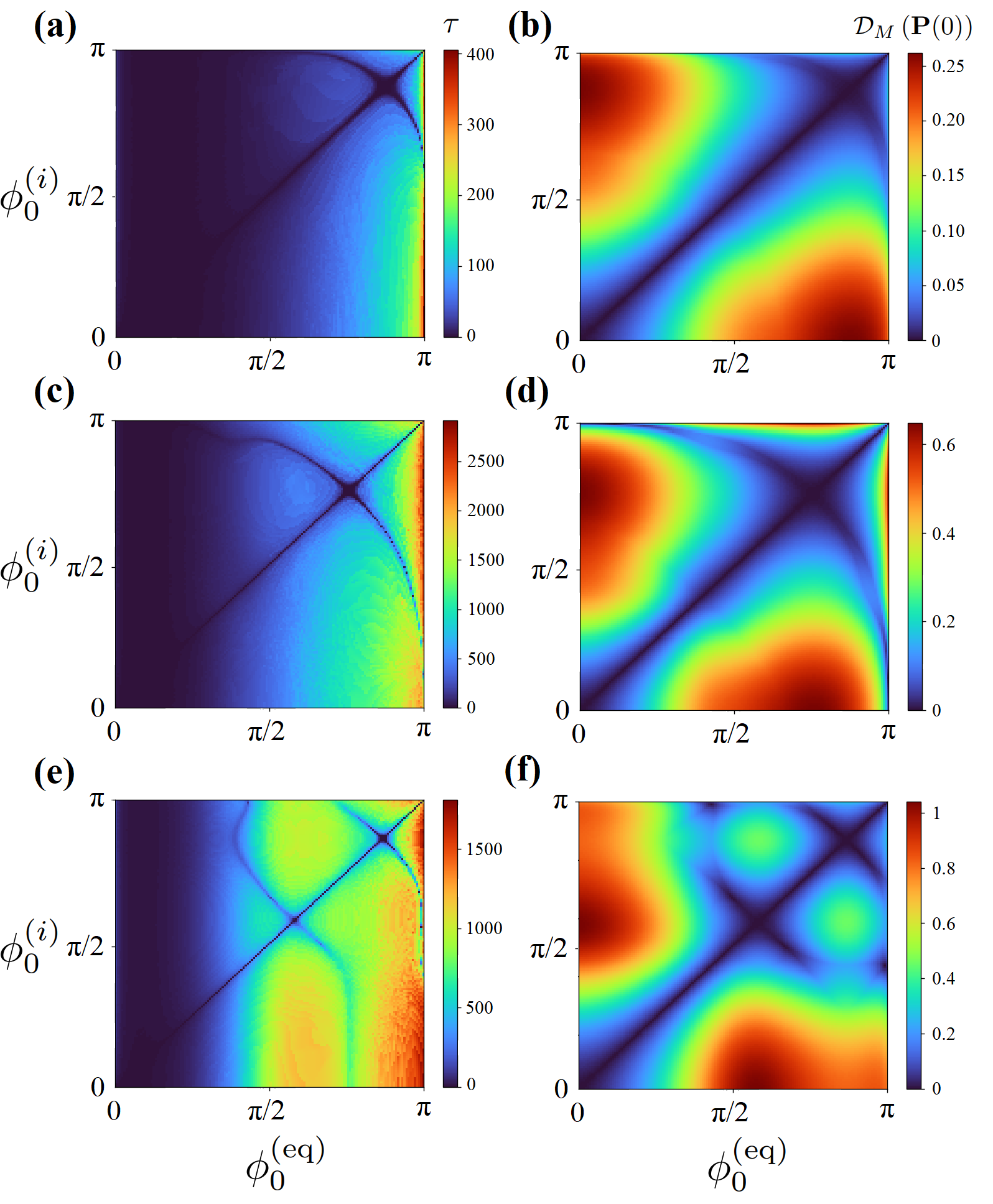}
    \caption{QME in Josephson dots of short length, $L=0.6\xi_0$. Results follow by numerical calculations for the model in Sec.~\ref{sec2} by computing $\mathcal{D}_{M}\left(\mathbf{P}(t)\right)$ in Eq.~\eqref{norm1}  from the Pauli equation.  Color-scale plots show  
    the relaxation time $\tau$ or $\mathcal{D}_{M}\left(\mathbf{P}(t=0 )\right)$
    in the plane spanned by pre-quench $(\phi_0^{(i=c,f)})$ and post-quench $(\phi_0^{({\rm eq})})$ phase differences.
    Panels (a), (c), and (e) show $\tau$ in units of $\Delta^{-1}$.  
    Panels (b), (d), and (f) show $\mathcal{D}_{M}\left(\mathbf{P}(0)\right)$.  With $\Delta=1$, we    
 use $T_b=0.2$, $T_{\rm qp}=0.5$, $\Omega_e=0.01$, $\eta=0.1$, $\kappa=0.1$ and $\Gamma=6.5$. 
 The junction transparency is ${\cal T}=0.3$ (with $\epsilon=9.8$) for panels (a) and (b); ${\cal T}=0.85$ (with $\epsilon=2.74$) in panels (c) and (d);
 and ${\cal T}=0.99$ (with $\epsilon=0.46$) in panels (e) and (f).
    }
    \label{fig5}
\end{center}
\end{figure}

\subsubsection{Arbitrary $\Gamma/\Delta$}\label{sec3a3}

If the condition $\Gamma\gg  \Delta$ is not satisfied, instead of Eq.~\eqref{ABS1}, one needs to employ the numerically exact ABS dispersion obtained by solving Eq.~(\ref{eom:ALScrheq}). Similarly, 
one then has to numerically evaluate the transition rates in Eq.~\eqref{pauli} from
the corresponding general expressions  in Sec.~\ref{sec2c}.  We have determined the relaxation times $\tau^{(i=c,f)}$ by computing the respective time-dependent distance functions $\mathcal{D}_{M}\left(\mathbf{P}(t)\right)$ in Eq.~\eqref{norm1} from the Pauli equation, using again the quench protocol for $\phi_0$ described above.  
Below, we also address what happens upon lowering the junction transparency ${\cal T}$.

In Fig.~\ref{fig5}, we show color-scale plots for $\tau$ and $\mathcal{D}_{M}\left(\mathbf{P}(0)\right)$ in the plane spanned by pre-quench ($\phi_0^{(i=c,f)}$) and post-quench $(\phi_0^{({\rm eq})})$ phases, using
three different values for ${\cal T}$.  
In each panel,  along the diagonal line $\phi_0^{(i)}=\phi_0^{({\rm eq})}$, 
no quench takes place, and therefore $\tau=\mathcal{D}_{M}\left(\mathbf{P}(0)\right)=0$.  
In the absence of a QME, the relaxation time $\tau$ is then expected to monotonically increase when 
moving away from this diagonal line at fixed $\phi_0^{({\rm eq})}$. 
However, from panels (a), (c), and (e) of Fig.~\ref{fig5}, 
we find that $\tau$ suddenly decreases again near special phase values 
$\phi_0^{(i)}=\phi_0^{(i,\ast)}\neq\phi_0^{({\rm eq})}$.  
At those values, $\mathbf{P}_{r,{\rm stat}} (\phi_0^{(i,\ast)})\approx \mathbf{P}_{r,{\rm stat}}(\phi_0^{({\rm eq})})$ for the stationary occupation probabilities, see  Eq.~\eqref{prt}.
The resulting functions $\phi_0^{(i,\ast)}(\phi_0^{({\rm eq})})$ define \emph{open} curves in the $\phi_0^{(i)}$-$\phi_0^{({\rm eq})}$ plane which cross
the main diagonal.  Below, we refer to such curves as ``Mpemba arcs.''  

In fact, the existence of at least one Mpemba arc is a necessary and sufficient condition for the QME to occur in this system.   Indeed, for any value of $\phi_0^{({\rm  eq})}$
such that $\phi_0^{(i,\ast)}$ exists,  a non-monotonic function $\tau(\phi_0^{(i)})$ is defined by the relaxation time which has a maximum at some value $\phi_0^{(M)}\in \left(\phi_0^{({\rm eq})},\phi_0^{(i,\ast)}\right)$.
It is then always possible to prepare two system copies in initial stationary states
corresponding to pre-quench values $\phi_0^{(c)}$ and $\phi_0^{(f)}$ 
such that Eq.~\eqref{condition_mpemba} is satisfied. For example, one can choose $\phi_0^{({\rm  eq})}\lessgtr \phi_0^{(M)} \lessgtr \phi_0^{(c)}\lessgtr \phi_0^{(f)}\lessgtr \phi_0^{(i,\ast)}$.

While analyzing the relaxation time $\tau$ in the $\phi_0^{(i)}$-$\phi_0^{({\rm eq})}$ plane is sufficient for establishing the QME, 
it is necessary to study $\mathcal{D}_{M}\left(\mathbf{P}(0)\right)$ in order to distinguish type-I and type-II QMEs.
A useful property for identifying the two types of QME follows from Eqs.~\eqref{norm1} and \eqref{pauli_sol}. Indeed,  $\mathcal{D}_{M}\left(\mathbf{P}(0)\right)$ is symmetric with respect to an exchange of the pre- and post-quench stationary states, and hence panels (b), (d), and (f) of Fig.~\ref{fig5} exhibit a mirror symmetry with respect to the main diagonal. 
However, since the decay rates $|\lambda_k|$ only depend on $\phi_0^{({\rm eq})}$, the relaxation times $\tau$ extracted from Eq.~\eqref{pauli_sol}, see panels (a), (c), and 
(e) of Fig.~\ref{fig5}, evidently do not display this mirror symmetry. 
As a consequence, a mismatch between the $\tau$-isolines and the $\mathcal{D}_{M}\left(\mathbf{P}(0)\right)$-isolines
in the $\phi_0^{(i)}$-$\phi_0^{({\rm eq)}}$ plane can occur.  Taken as a function of $\phi_0^{(i)}$ at fixed $\phi_0^{({\rm eq})}$, the minima of $\tau$ 
may therefore be located at different positions than the minima of $\mathcal{D}_{M}\left(\mathbf{P}(0)\right)$.
In such cases, Eq.~\eqref{condition_mpembaII} can be fulfilled, resulting in a type-II QME \cite{Nava2024}.  

By varying the bare dot level energy $\epsilon$, e.g., by means of a gate voltage applied on the dot, one may change the junction transparency ${\cal T}$, see Eq.~\eqref{transparency_short}. 
From panels (a), (c), and (e) of Fig.~\ref{fig5}, we observe that Mpemba arcs tend to move towards higher values of $\phi_0$ with decreasing ${\cal T}$, 
eventually fading away for small transparency.
As in the case of large $\Gamma/\Delta$, see Eq.~\eqref{ABS1}, we find that for highly transparent contacts (${\cal T}\approx 1$), the ABS dispersion essentially spans the full subgap energy range. 
For very small ${\cal T}$, on the other hand, the ABS energy dispersion approaches the continuum threshold from below, i.e., $E_1(\phi_0)\approx \Delta$ for all $\phi_0$. 
As shown in Figs.~\ref{fig3} and \ref{fig4}, extrema in the stationary populations as a function of $\phi_0$ are typically obtained for intermediate values of $E_1(\phi_0)$ within the gap. 
As discussed above, at such extremal points, the competition between all transition rates in Eq.~\eqref{pauli} becomes crucial. 
For small transparency, we find that Mpemba arcs tend to disappear or fade away, see Fig.~\ref{fig5}(a),  since now the rate $\Gamma_a^{-}$ dominates over all other transition 
rates for the realizable values of $E_1(\phi_0)$. 
With increasing ${\cal T}$, see Fig.~\ref{fig5}(c,e), the ABS energy $E_1(\phi_0)$ can reach lower values, and ratios between transition rates can change from $<1$ to $>1$ as function of $\phi_0$.
As a consequence, one or several well-defined sharp Mpemba arcs can emerge.  Since the mirror reflection asymmetry of the relaxation time is then also more pronounced, a
type-II QME is commonly encountered. This conclusion is qualitatively consistent with our results for high transparency in Fig.~\ref{fig3} and \ref{fig4}. 

\begin{figure}
\begin{center}
\includegraphics[width=0.45\textwidth]{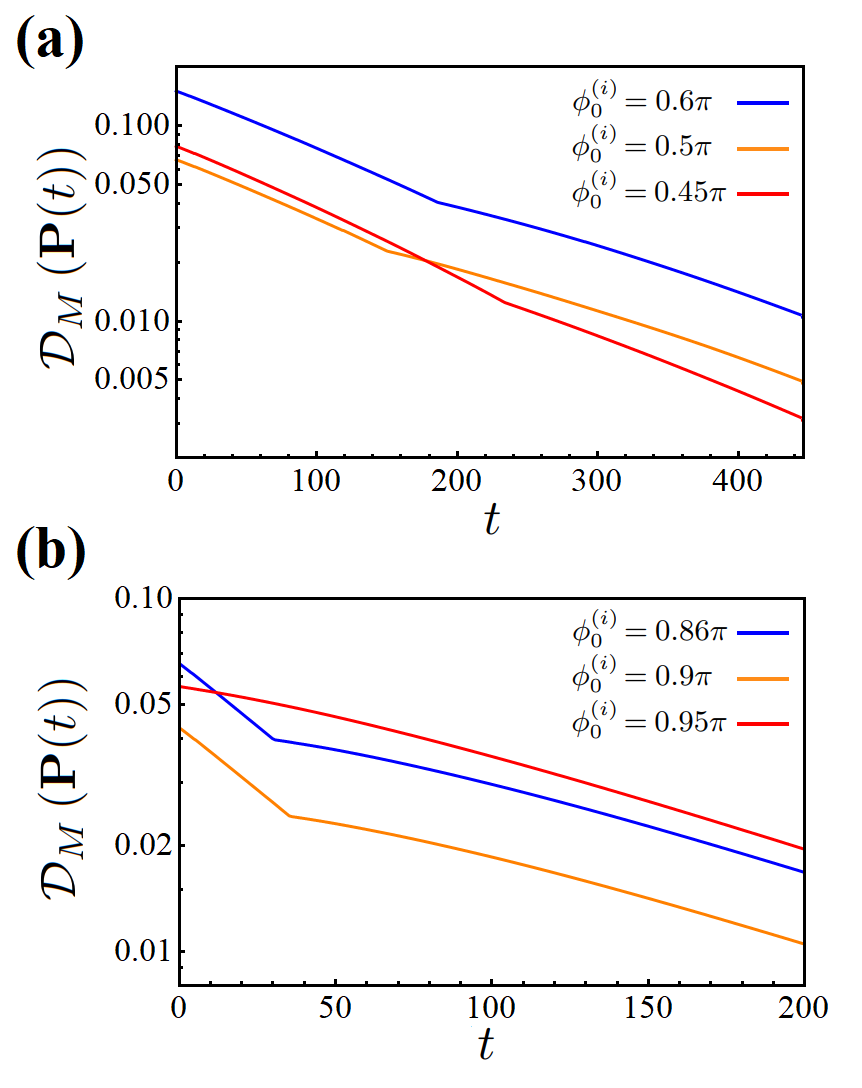}
\end{center}
\caption{Time dependence of the distance function $\mathcal{D}_{M}\left(\mathbf{P}(t)\right)$, see Eq.~\eqref{norm1}, for a short Josephson dot with the parameters in Fig.~\ref{fig5}(e). Time is given in units of $\Delta^{-1}$. Note the semi-logarithmic scales.
In panel (a), we consider $\phi_0^{({\rm eq})}=0.7\pi$ and $\phi_0^{(i)}=0.6\pi$ (blue curve), $\phi_0^{(i)}=0.5\pi$ (orange), and $\phi_0^{(i)}=0.45\pi$ (red). In panel (b), $\phi_0^{({\rm eq})}=0.8\pi$ with $\phi_0^{(i)}=0.86\pi$ (blue), $\phi_0^{(i)}=0.9\pi$ (orange), and $\phi_0^{(i)}=0.95\pi$ (red).}
\label{fig6}
\end{figure}

 In Fig.~\ref{fig6}, we show $\mathcal{D}_{M}\left(\mathbf{P}(t)\right)$ for selected values of $\phi_0^{(i)}$ and $\phi_0^{({\rm eq})}$, using the same parameters as in Fig.~\ref{fig5}(e) 
for a highly transparent junction with ${\cal T}=0.99$.
Let us start with Fig.~\ref{fig6}(a), where the blue (red) curve corresponds to the closest (farthest) initial configuration $\phi_0^{(i)}$ 
with respect to the final state defined by $\phi_0^{({\rm eq})}=0.7\pi$. The orange curve $(\phi_0^{(i)}=0.5\pi)$ represents 
an intermediate situation. 
We first observe that for $\phi_0^{(c)}=0.6\pi$ with $\phi_0^{(f)}=0.45\pi$ or $\phi_0^{(f)}=0.5\pi$, a type-I QME emerges.
Second, a type-II QME takes place when choosing $\phi_0^{(c)}=0.5\pi$ and $\phi_0^{(f)}=0.45\pi$, since the time evolution of the distance function now exhibits
a crossing point at $t^\ast \approx 180\Delta^{-1}$. 
It is worth noting that all curves have the same slope at long times since the system relaxes to the stationary state with the same rate $|\lambda_1|$.

Next, in Fig.~\ref{fig6}(b), we investigate the case  $\phi_0^{({\rm eq})}=0.8\pi$. As in Fig.~\ref{fig6}(a), the color assignment is such that the blue (red) curve corresponds to the 
closest (farthest) initial condition and the orange curve represents an intermediate case. A type-I QME can be observed by choosing $\phi_0^{(c)}= 0.86\pi$ (blue curve) and $\phi_0^{(f)}=0.9\pi$ (orange), while no QME take places for $\phi_0^{(c)}= 0.9\pi$ and $\phi_0^{(f)}=0.95\pi$ (red).
We identify the quench dynamics defined by the blue and red curves in Fig.~\ref{fig6}(b) as an ``avoided'' QME rather than a type-II QME. Indeed,
even though a crossing point exists at time $t^\ast\approx 15\Delta^{-1}$, the blue curve (corresponding to the ``close'' initial condition)  describes a faster relaxation at
long times despite the fact that $\mathcal{D}_{M}\left(\mathbf{P}^{(f)}(0)\right)<\mathcal{D}_{M}\left(\mathbf{P}^{(c)}(0)\right)$.  However, since Eq.~\eqref{condition_mpemba} does not hold, no QME emerges according to our definitions.  
We note that avoided QMEs have previously been classified as QMEs in the literature
due the existence of a crossing point in the monitoring function, see, e.g., Refs.~\cite{Wang2024,Joshi2024,Rylands2023}.

\subsection{Intermediate-length junction with SOI and Zeeman field}\label{sec3b}

\begin{figure}
\begin{center}
    \includegraphics[width=0.42\textwidth]{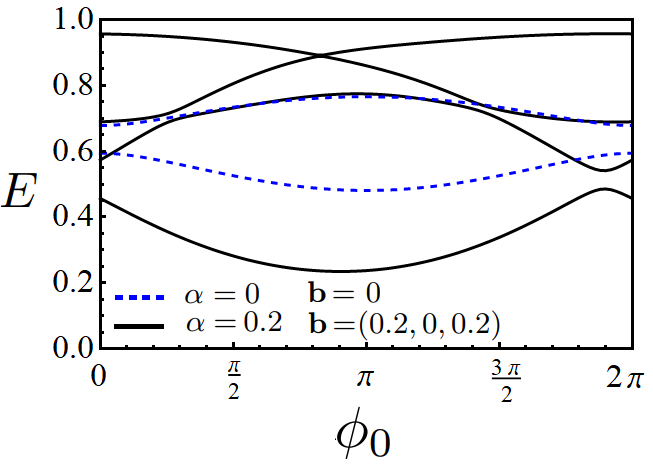}
    \caption{The four positive ABS energies vs $\phi_0$ 
    for a Josephson dot of intermediate length, $L=1.7\xi_0$, 
    in a Zeeman field $\mathbf{b}=\left(0.2,0,0.2\right)$ with SOI coupling $\alpha=0.2$ (black solid curves), where we set $\Delta=1$. The presence of both SOI and Zeeman field breaks both the spin symmetry (i.e., each blue curve represents a spin degenerate ABS energy) and the $\phi_0 \rightarrow 2\pi-\phi_0$ symmetry so that $E_\lambda(2\pi-\phi_0)\neq E_\lambda(\phi_0)$ \cite{Zatsarynna2024}. The case $\alpha=0$ and ${\bf b}=0$ is shown by the blue dashed curves for comparison. }
    \label{fig7}
\end{center}
\end{figure}

In this section, we investigate the QME for an intermediate-length junction. Increasing the length of the nanowire, the low-energy transport properties are described by more 
single-particle eigenstates of $H_{\rm dot}$. For instance, for a weak link of
intermediate length $L \approx \xi_0=v_{\rm F}/\Delta$, typically four single-particle ABSs are found \cite{Ackermann2023,Zatsarynna2024}. 
If both SOI and Zeeman are present, orbital and spin angular momenta are no longer conserved,
and the (positive) ABS energies split into four distinct levels. Ordering these 
energies $E_\lambda(\phi_0)$ by increasing energy, $0< E_1 \le E_2 \le E_3 \le E_4 < \Delta$,
see Fig.~\ref{fig7} for our numerical results for their energy dispersions, we have 16 many-body Andreev states.
We label these many-body states as $\ket{n_1,n_2,n_3,n_4}$ with $n_\lambda \in \left\{0,1 \right\}$, where $n_\lambda=0$ ($n_\lambda=1$) means that the corresponding ABS level is unoccupied (occupied). The dynamics for the occupation probabilities is still described by the Pauli master equation in Eq.~\eqref{pauli_master}, where now $\mathbf{M}$ is a $16\times 16$ matrix \cite{Ackermann2023,Zatsarynna2024}. 

\begin{figure}
\begin{center}
    \includegraphics[width=0.5\textwidth]{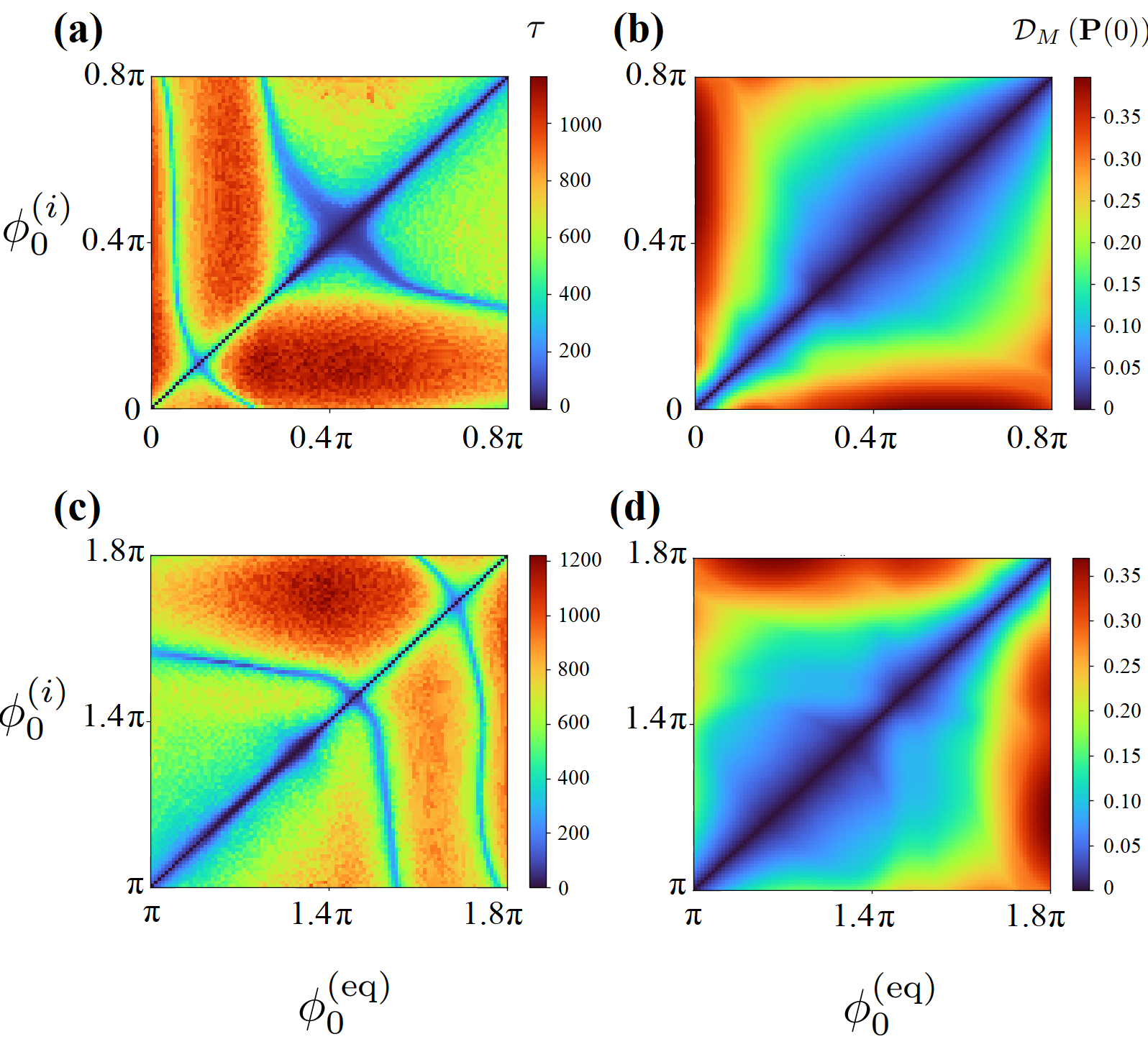}
    \caption{QME in a Josephson dot of intermediate length, $L=1.7\xi_0$,
    for a Zeeman field $\mathbf{b}=\left(0.2,0,0.2\right)$ and SOI coupling $\alpha=0.2$, where we set $\Delta=1$. 
    Results are obtained from the Pauli equation for the model in Sec.~\ref{sec2} by monitoring the distance function $\mathcal{D}_{M}\left(\mathbf{P}(t)\right)$ in Eq.~\eqref{norm1} after the quench. Color-scale plots show  the relaxation time $\tau$, see panels (a) and (c), and the distance function at time $t=0$, see panels (b) and (d), in a plane spanned by the pre-quench phase, $\phi_0^{(i)}$, and the post-quench value $\phi_0^{({\rm eq})}$.
    The hopping parameters in Eq.~\eqref{md:Htunx} are $t_1=t_2=0.7$, with $m_x=9/v_{\rm F}^2$ in Eq.~\eqref{md:Hdot}, where the four dot energy levels in Eq.~\eqref{md:Hdotc},
    $\epsilon_\nu \in \{-0.1,  0.41,  0.62,  0.97\}$, correspond to a high-transparency junction.
    The other parameters are $T_b=0.2$, $T_{\rm qp}=0.5$, $\Omega_e=0.01$, $\eta=0.1$, and $\kappa=0.1$.}
    \label{fig8}
\end{center}
\end{figure}

Even though the system dynamics is now considerably richer, the qualitative effect of the impact of more ABS levels and/or the inclusion of SOI and Zeeman field effects on the QME can be assessed again by starting from a simplified argument as in Sec.~\ref{sec3a1}.  Indeed, for each ABS energy, the same approximation as employed in Sec.~\ref{sec3a1} can be applied separately.
If $E_\lambda$ approaches $\Delta$, the transition of an ABS quasiparticle to the continuum or vice versa, see Fig.~\ref{fig2}(a,d), is the dominant process. On the other hand, Cooper pair processes without contributions of a continuum quasiparticle, see Fig.~\ref{fig2}(c,f), dominate for $E_\lambda \ll \Delta$. If a continuum quasiparticle is involved in a pair process, see Fig.~\ref{fig2}(b,c), the corresponding rate is always subleading.  However, for longer junctions, additional processes (absent in short junctions) appear since transitions of quasiparticles between different ABS levels become now possible, with the corresponding transition rates in Eq.~\eqref{as:Glmbdn}. These processes preserve the occupation number of the ABS sector and dominate if both ABS energies are close, $E_\lambda \approx E_{\lambda'}$, regardless of their position within 
the superconducting gap. Such processes are especially important if both ABS energies are 
near $\approx \Delta/2$ since the other processes listed above then become ineffective.

After the quantum quench, the occupation probabilities of the many-body Andreev states relax to
their final steady-state values, which have a characteristic $\phi_0$-dependence 
as shown in Fig.~\ref{fig3} for a short junction.  With increasing junction length,
features like common extremal points of different occupation probabilities as function of $\phi_0$, see Fig.~\ref{fig3}(c), become less and less likely as the number of ABSs increases. Qualitatively, we thus expect that by increasing $L$, and allowing for finite SOI and Zeeman field, the QME is either washed out or, if it survives, the type-II QME becomes more common than the type-I QME.  We next check this prediction using numerical calculations.

In Fig.~\ref{fig8}, we show color-scale plots of $\tau$ and $\mathcal{D}_{M}\left(\mathbf{P}(0)\right)$ in a plane spanned by the pre-quench ($\phi_0^{(i)}$) and post-quench $(\phi_0^{({\rm eq})})$ phase differences for an intermediate-length junction with $L=1.7 \xi_0$ in the presence of SOI and a Zeeman field.
 The remaining parameters are chosen as in Fig.~\ref{fig5}, i.e., they correspond to a QME regime in the short-junction case. 
 Due the presence of the magnetic field and the SOI, the mirror symmetry of the ABS spectrum is broken, $E_\lambda(2\pi-\phi_0)\neq E_\lambda(\phi_0)$. For this reason, in order to avoid ambiguities in the quench protocol, 
 we restrict the phase differences $\phi_0$ studied below to a range in which all ABS energies $E_\lambda(\phi_0)$ are monotonic functions, see Fig.~\ref{fig7}. In  Fig.~\ref{fig8}(a,b), we
 show the relaxation time $\tau$ and $\mathcal{D}_{M}\left(\mathbf{P}(0)\right)$ by 
 assuming that both $\phi_0^{({i})}$ and $\phi_0^{({\rm eq})}$ vary in the interval $\left[0, 0.8\pi\right]$.  
In Fig.~\ref{fig8}(c,d), these phases instead belong to the interval $\left[\pi, 1.8\pi\right]$. 

Figure~\ref{fig8}(a,c) reveals the presence of Mpemba arcs once again. This is a signature that QMEs are 
still realizable in the longer junction considered here.
However, since the mirror symmetry under $\phi_0 \rightarrow 2\pi-\phi_0$ is broken by the interplay of SOI and Zeeman field, the Mpemba arcs are not symmetric under point reflections with respect to $(\pi,\pi)$.
Furthermore, in contrast to the short-junction case in Fig.~\ref{fig5}, we note that Mpemba arcs are not visible  in $\mathcal{D}_{M}\left(\mathbf{P}(0)\right)$ anymore, see Fig.~\ref{fig8}(b,d). As a consequence, if Eq.~\eqref{condition_mpemba} is satisfied, also Eq.~\eqref{condition_mpembaII} must be true. 
We conclude that the more elusive type-II QME is therefore the dominant type of QME realized 
for a longer Josephson dot in the presence of SOI and Zeeman field.

\section{Conclusions}\label{sec4}

In this work, we have introduced a general GF-based framework for studying the dynamics of quasiparticles in 
multi-level quantum dot systems coupled to multiple reservoirs.  The formalism has been described in detail for
the case of two superconducting leads, where we allow for SOI and Zeeman fields in the dot region defining the Josephson junction.  The general formalism has then been applied to a study of Mpemba effects in this open quantum system context.
Following the protocol proposed in Ref.~\cite{Nava2024}, we have shown that already the simplest case of a short 
junction harbors both allowed types of QME which can be observed in phase quench experiments.  To that end, one has
to monitor the distance function for the occupation probabilities in Eq.~\eqref{norm1}, which in turn are experimentally 
accessible by microwave spectroscopy \cite{Janvier2015,Wesdorp2023,Wesdorp2024}.   The setup studied in this paper thus offers a simple platform where quantum generalizations of the Mpemba effect can be systematically studied.
We hope that our paper will stimulate further theoretical and experimental work along these lines.

\new{All data underlying the figures presented in this paper can be retrieved at zenodo \cite{Zatsarynna_2025_zenodo}.}

\begin{acknowledgments}
  We acknowledge funding by the Deutsche Forschungsgemeinschaft (DFG, German Research Foundation) Grant No.~277101999 - TRR 183 (project C01) and under Germany's Excellence Strategy - Cluster of Excellence Matter and Light for Quantum Computing (ML4Q) EXC 2004/1 - 390534769.  
\end{acknowledgments}

\begin{appendix}
\section*{Appendix: Dot wave functions}
\setcounter{equation}{0}

\renewcommand{\theequation}{A\arabic{equation}}

We here provide details concerning the calculation of the wave functions $\chi_\nu(x)$ of an isolated quantum dot in the presence of SOI and Zeeman field, Eq.~\eqref{doteigen}, which determine the hybridization matrices \eqref{hybrid}. Without loss of generality, we choose a coordinate system where the wire ends $x_{1,2}$ are located at $x_1 = - x_2 = -L/2$, and we assume $V(x) = 0$ inside the wire region $x \in (x_1, x_2)$. First, we solve the eigenvalue problem Eq.~\eqref{doteigen} in the absence of Zeeman splitting, i.e., for ${\bf b} = 0$. In this case, the spin-up and spin-down eigenstates of the Hamiltonian \eqref{md:hath} are decoupled, and the spin bands minima are shifted by $q = m_x \alpha$ in opposite directions along the momentum ($\hat p$) axis. Applying the Neumann boundary conditions, $\partial_x \chi(-L/2) = \partial_x \chi(L/2) = 0$, the complete set of orthonormal eigenfunctions $\chi_{n \sigma}(x)$ for spin projection $\sigma$ is given by
\begin{eqnarray}\label{app:chin}
\chi_{n \sigma}(x) &=& \frac{e^{-i \sigma q x}}{\sqrt{L}} \Bigl[
\cos (\theta_{n \sigma}) e^{i k_n (x - L/2)} \\ \nonumber && \qquad  + \sin( \theta_{n \sigma}) e^{-i k_n (x - L/2)}
\Bigr],
\end{eqnarray}
with the corresponding eigenenergies $\varepsilon_n = k_n^2/(2m_x) - m_x \alpha^2 / 2 - \mu$, where
\begin{equation}
k_n = \pi n / L,\quad \cos \theta_{n \sigma} = \sin \theta_{n, -\sigma} =
\frac{k_n + \sigma q}{\sqrt{2 \left( k_n^2 + q^2 \right)}},
\end{equation}
and $n \in \{ 1, 2, ... \}$. Some comments are in order at this point. (i) In the absence of the Zeeman field, we have an infinite set of 
spin-degenerate dot levels $\varepsilon_n$. (ii) Each state $\chi_{n \sigma}(x)$ is a linear combination of left and right movers. (iii) While spatial inversion symmetry is broken for $\alpha \neq 0$, time-reversal symmetry is present as reflected by the relation
$\chi_{n \sigma}^\ast(x) = \chi_{n, -\sigma}(x).$
(iv) Finally,  boundary values of  wave functions are connected by the relation
$\chi_{n \sigma}(-L/2) = (-1)^n e^{i \sigma q L} \chi_{n \sigma}(L/2).$
We note that the phase shift due to $e^{i \sigma q L}$ in this relation does not affect electronic transport  since it can be removed from the total Hamiltonian 
 by a gauge transformation on the lead fermions.

Next, we add the Zeeman field ${\bf b} = (b_x, 0, b_z)$ to our consideration and choose the wave functions \eqref{app:chin} as a basis set for constructing the dot-level representation of the full eigenvalue problem. Any solution of Eq.~\eqref{doteigen} that satisfies the Neumann boundary conditions can then be written as
\begin{equation}
\chi_\nu(x) = \sum_{n=1}^\infty \left(
\begin{array}{c} u_{n \uparrow}^{(\nu)} \chi_{n \uparrow}(x) \\
u_{n \downarrow}^{(\nu)} \chi_{n \downarrow}(x) \end{array} \right),
\end{equation}
with complex-valued amplitudes $u_{n \sigma}^{(\nu)}$ corresponding to the energy eigenvalue $\epsilon_\nu$. Introducing the multispinor
\begin{equation}
\Psi_\nu = \left( u_{1 \uparrow}^{(\nu)}, u_{1 \downarrow}^{(\nu)}, u_{2 \uparrow}^{(\nu)}, u_{2 \downarrow}^{(\nu)}, ... \right)^T,
\end{equation}
we can write Eq.~\eqref{doteigen} as
\begin{equation}
{\cal H} \Psi_\nu = \epsilon_\nu \Psi_\nu,
\end{equation}
with an infinite-size matrix
\begin{equation}\label{app:calH}
{\cal H} = \left( \begin{array}{ccccc} \varepsilon_1 + b_z &  J_{11}^- &
0 & J_{12}^- &  ... \\
J_{11}^+ & \varepsilon_1 - b_z &  J_{12}^+ &
0 &   \\
0 & J_{21}^- & \varepsilon_2 + b_z & J_{22}^- &   \\
J_{21}^+ & 0 & J_{22}^+ & \varepsilon_2 - b_z &  \\
... &&&& ...
\end{array} \right),
\end{equation}
where
\begin{equation}\label{app:Jmn}
J_{mn}^\sigma = \left( J_{n m}^{-\sigma} \right)^\ast = b_x \int_{-L/2}^{L/2} dx \, \chi^\ast_{m,-\sigma}(x)\chi_{n \sigma}(x).
\end{equation}
Explicitly, the overlap integrals \eqref{app:Jmn} are given by
\begin{equation}
J_{mn}^\sigma = - \frac{2b_x}{L}f^\sigma_{mn} \times
\left\{
\begin{array}{cl} \sigma \sin(qL) & {\rm for~} (-1)^{n+m} = 1 \\
i \cos(qL) & {\rm for~} (-1)^{n+m} = -1 \end{array}
\right.,
\end{equation}
where
\begin{eqnarray}
  f^\sigma_{mn} & = & \frac{\cos \theta_{n \sigma} \cos \theta_{m,-\sigma}}{k_n - k_m - 2 \sigma q} -
\frac{\sin \theta_{n \sigma} \sin \theta_{m,-\sigma}}{k_n - k_m + 2 \sigma q} 
\nonumber \\ 
 &+& \frac{\cos \theta_{n \sigma} \sin \theta_{m,-\sigma}}{k_n + k_m - 2 \sigma q} -
\frac{\sin \theta_{n \sigma} \cos \theta_{m,-\sigma}}{k_n + k_m + 2 \sigma q},
\end{eqnarray}
with the   symmetry properties
\begin{equation}
f^\sigma_{mn} = f^\sigma_{nm} = \left( f^\sigma_{nm} \right)^\ast,\quad
f^{-\sigma}_{mn} = - f^\sigma_{mn}.
\end{equation}
In practice, the Hamiltonian matrix \eqref{app:calH} is truncated to the subspace of low-energy eigenstates below some energy cutoff comparable to $\Delta$, and then diagonalized numerically. 
\end{appendix}

\bibliography{biblio}
\end{document}